\documentclass[aps,nofootinbib,twocolumn,preprintnumbers,superscriptaddress,nobibnotes,floatfix,longbibliography]{revtex4-1}
\usepackage{graphicx}
\usepackage[utf8]{inputenc}
\usepackage{amsmath}
\usepackage{amssymb}

\usepackage{array}
\usepackage{slashed}

\usepackage{color,soul}
\usepackage{booktabs}
\usepackage{mathtools}
\usepackage{xfrac}
\usepackage{xcolor}
\definecolor{light-purple}{RGB}{184,82,255}
\definecolor{pale-orange}{HTML}{FF9300}
\usepackage{mdframed}

\providecommand{\main}{.}
\graphicspath{{images/}{\main/images/}}

\usepackage{nameref}

\usepackage[a4paper, total={7in, 10in}]{geometry}

\usepackage{orcidlink}         
\usepackage{hyperref}

\begin{document}
\setcounter{secnumdepth}{3}
\title{Linking Leptogenesis and Asymmetric Dark Matter: A Testable Framework for Neutrino Mass and the Matter-Antimatter Asymmetry}

\author{Henry McKenna\,\orcidlink{0009-0009-1474-0011}}
\email{Henry.McKenna@liverpool.ac.uk}
\affiliation{Department of Mathematical Sciences, University of Liverpool, Liverpool, L69 7ZL, United Kingdom}

\author{Juri Smirnov\,\orcidlink{0000-0002-3082-0929}}
\email{juri.smirnov@liverpool.ac.uk}
\affiliation{Department of Mathematical Sciences, University of Liverpool, Liverpool, L69 7ZL, United Kingdom}

\author{Martin Gorbahn\,\orcidlink{0000-0002-8863-7318}}
\email{mgorbahn@liverpool.ac.uk}
\affiliation{Department of Mathematical Sciences, University of Liverpool, Liverpool, L69 7ZL, United Kingdom}

\begin{abstract}
We investigate a minimal extension of the Leptogenesis framework that simultaneously explains the observed baryon asymmetry and dark matter (DM) abundance through the decay of a heavy Majorana neutrino. In this scenario, CP violation arises from complex Yukawa couplings, enabling the generation of asymmetries in both the Standard Model (SM) and DM sectors. We explore two regimes: (i) wash-in, where an initial dark asymmetry is transferred to SM leptons by $2 \leftrightarrow 2$ scattering processes; and (ii) co-genesis, featuring a hierarchical coupling structure that allows enhanced CP violation while supporting a low-scale seesaw mechanism at order $\mathcal{O}(2)$ TeV. This setup not only links light neutrino masses to the Majorana mass term but also suggests that lepton-number violation may occur at experimentally accessible energy scales. In the co-genesis scenario, we show spin-independent cross sections for DM heavier than 10 GeV that can be tested in current direct detection experiments and motivate the exploration of cross sections inside the neutrino fog for lighter DM masses, establishing asymmetric leptogenesis as a predictive benchmark framework for direct-detection experiments and identifying a new hierarchical-coupling regime enabling TeV-scale leptogenesis.
\end{abstract}
\maketitle

\section{Introduction}

Matter and dark matter provide a substantial fraction of the energy budget of our universe. Although we do not know what microscopic interactions govern the dark-matter sector, we know that interaction rates, and annihilation rates of known Standard Model (SM) fields are large. Thus, given fast annihilation in the early universe, the abundance of symmetric SM particles is expected to be tiny, which is compatible with cosmological observations. At the same time, we observe an asymmetry in the SM sector that leads to a relative overabundance of matter over antimatter fields of the order of one part per billion, or more accurately 
\cite{WMAP:2010qai} 

\begin{equation}
    \eta_B = \frac{n_B}{n_\gamma} = (6.19 \pm 0.15)\times 10^{-10},
\end{equation}
where $n_B$ is the number density of baryonic matter and $n_\gamma$ is the photon number density of our universe. This indicates that the charge-parity (CP) symmetry in the SM sector must be violated at a yet unknown energy scale. 

For the dark matter (DM) sector it is unknown which fields and symmetries govern its dynamics, and whether self-annihilations deplete its symmetric abundance similarly to the SM fields. In scenarios where the symmetric DM abundance freezes out at exactly the required DM density, we speak of thermal production, which can be facilitated by many types of interactions leading to a variety of DM interaction scenarios, see Refs.~\cite{Arcadi:2017kky,Bertone:2004pz,Pospelov:2007mp,Griest:1990kh} for a non-exhaustive list of freeze-out scenarios. However, in this work we make the assumption that the symmetric part of the DM has annihilated away sufficiently, similarly to the SM, and the DM abundance is set by a non-thermal process that violates the CP-symmetry in the dark sector~\cite{Petraki:2013wwa, An:2009vq}. Similarly to Ref.~\cite{Falkowski:2011xh} we assume that this happens in a minimal scenario, where CP-violation is induced by decays of a heavy Majorana neutrino. 
\\
This scenario is a minimal extension of the well-known leptogenesis mechanism, where the three Sakharov conditions, namely the departure from thermal equilibrium, C and CP-violation, and baryon-number violation, are satisfied by the decay of a heavy Majorana neutrino. 
This particle has a sufficiently small decay rate to temporarily decouple from its thermal distribution in the early universe. It has, furthermore, complex Yukawa couplings, which due to Feynman diagram interference lead to CP violation in its decays to SM leptons and DM particles. Non-perturbative SM processes, known as sphalerons, lead to baryon plus lepton ($B+L$) number violation, while preserving the $B-L$ symmetry. Thus, satisfying the Sakharov condition \cite{Sakharov:1967dj} by transferring the generated lepton asymmetry to the baryon sector, see Refs.~\cite{Bodeker:2020ghk, Davidson:2008bu} for a recent review, as well as Refs.~\cite{Moffat:2018wke,Brivio:2019hrj,Fu:2022lrn,Baker:2024udu} for additional specific examples.
\\
In this work, we consider two previously overlooked regimes of this minimal scenario that provide the DM and SM abundances by the same mechanism resulting in Asymmetric Dark Matter \cite{Kaplan:2009ag,Petraki:2013wwa,Chand:2025clg,Zurek:2013wia,Agrawal:2016uwf, Datta:2025vyu, Kanemura:2025ixp, Dessert:2018khu}. One is the possibility that the matter-antimatter asymmetry is dominantly produced in the DM sector, and then transferred via $2 \leftrightarrow 2$ scattering processes to the SM leptons called wash-in \cite{Domcke:2020quw,Domcke:2022kfs,Schmitz:2023pfy,Asadi:2025vli}. The other is a possible coupling strength hierarchy that allows boosting of the CP-violating processes, while still maintaining sufficiently small Yukawa couplings of the Majorana neutrinos to SM fields, to allow a low-scale seesaw scenario; see, for example, \cite{Pilaftsis:2003gt,Alanne:2018brf,Croon:2022gwq, Spalding:2026jia}. 
\\
In our framework, we explore a minimal scenario in which the light neutrino masses, required to explain neutrino oscillations~\cite{Super-Kamiokande:1998kpq, SNO:2002tuh, KamLAND:2002uet, ParticleDataGroup:2024cfk}, are generated by mixing with Majorana neutrinos, the type-I seesaw~\cite{Cai:2017mow}. For a related framework with type-I seesaw in which right-handed neutrinos mediate the visible and dark sectors see~\cite{Li:2022bpp}. For a similar leptogenesis framework with type-III seesaw see \cite{Mahapatra:2023dbr}. At the same time the observed DM and SM abundance is linked to the fermion number violation induced by the Majorana mass term of the introduced sterile neutrinos. Unlike previous frameworks on ADM from Leptogenesis, we identify a new parametric regime that arises from both a large hierarchy in the dark sector couplings and a mild mass splitting between the RHN's still outside the resonant leptogenesis regime. Our novel finding is that this new mass scale associated with fermion (lepton) number violation can be significantly lower than expected in the standard seesaw scenario~\cite{Davidson:2002qv}. From the theoretical point of view this is appealing as a new heavy mass scale might not be required, between the electroweak and the Planck scale, which would alleviate the technical naturalness problem of why the observed Higgs boson mass is so close to the electroweak scale, despite not being protected from quantum corrections stemming from heavier fields \cite{Vissani:1997ys}. 
\\
Given the relatively low ($\sim 2 \, \text{TeV}$) mass scale associated with lepton-number violation, we explore how this can lead to detectable signals. We discuss the expected DM mass ranges, and its expected elastic scattering cross sections with Standard Model (SM) particles. The interactions are induced by particle loops involving the Majorana neutrino, or the scalar field $\phi$. Therefore, in theoretically motivated scenarios with lighter Majorana neutrinos the DM-SM elastic scattering cross sections are maximal, and can lead to detectable signals in direct detection experiments \cite{LZ:2024zvo,Battaglieri:2017aum,Essig:2011nj,Essig:2015cda,Ibe:2017yqa,Dolan:2017xbu}.

The present work introduces three key advances relative to previous asymmetric-dark-matter scenarios: (i) we identify a new hierarchical-coupling regime enabling successful leptogenesis at TeV-scale RHN masses without resonant degeneracy; (ii) we demonstrate that the parameters governing asymmetric production simultaneously determine direct-detection scattering rates, predicting correlated experimental targets; and (iii) we show that asymmetric dark matter from leptogenesis defines a production-driven benchmark framework for direct-detection searches, similar to thermal freeze-out benchmarks in WIMP models.

\section{The Dark Leptogenesis Framework}
In the following, we specify the field content, interactions, symmetries and assumptions underlying our model and give rise to the terms in the Boltzmann equations for the quantitative discussion of the two scenarios we consider.
The minimal model contains the SM fields, two heavy right-handed neutrinos (RHNs) $N_i$ (flavour index $i$), a singlet Dirac fermion $\chi$ and singlet complex scalar $\phi$.
Here, $\chi$ and $\phi$ constitute the dark sector, and the RHNs form the portal to the SM.
With lepton number only violated by the RHN Majorana mass, the relevant interactions are 
\begin{equation}
\label{eq:lag}
\begin{split}
   - \mathcal{L} \supset\,& \frac{1}{2} M_{N_i} \overline{N_i^c} N_i  + \lambda_{i} \,\overline{N}_i \chi \phi + 
   y_{\alpha i}\, \overline{N}_i L_{\alpha} \cdot H 
   \\& \ \ \ \ \ \ \ \ +  \frac{1}{2}\, \lambda_\phi \,\phi^\dagger \phi H^\dagger H+ h.c.,
\end{split}
\end{equation}
where  $N_i$, $\chi$, and $L_\alpha$ carry lepton number $+1$ and $L_\alpha$ denotes the left-handed lepton doublet with flavour index $\alpha \in \left\{ e,\mu,\tau \right\}$, $M_{N_i}$ is the Majorana mass of the particle $N_i$.
We assume the Dirac fermion $\chi$ (the DM candidate) and the complex singlet scalar $\phi$ are both charged under a $\mathbb{Z}_2$ symmetry. The DM has a mass term  $m_\chi \bar{\chi} \chi$ and the singlet scalar has a mass term $m_\phi^2\, \phi^\dagger \phi$. The model of Ref.~\cite{Falkowski:2011xh} corresponds to $\lambda_\phi=0$ with complex $\lambda_i$ and $y_{\alpha i}$. 
\\
A sphaleron conversion rate of $Y_{\Delta B}=12 Y_{\Delta L}/37$ \cite{Davidson:2008bu} is assumed since only the SM particles contribute to the sphaleron process. Standard-Model lepton flavour effects are not accounted for since we work in the unflavoured approximation with a single $Y_{\Delta L}$ and use the flavour-summed CP asymmetry $\epsilon_L=\sum_{\alpha}\epsilon_{L,\alpha}$, expressed via $(y^\dagger y)_{ij}=\sum_{\alpha} y^{\ast}_{\alpha i}y_{\alpha j}$. A full treatment including flavour effects may alter the asymmetry generation, but this is not expected to be significant enough to change the conclusions regarding the mass bound and detection outlook.
For simplicity, we define the square root of the diagonal terms as $|y_i|=\sqrt{(y^\dagger y)_{ii}}$.
\\
We explore two potential mechanisms for leptogenesis with asymmetric dark matter. The first scenario has a vanishing $|y_1|$ coupling, such that the right-handed neutrino decays lead to purely dark asymmetry generation, that is sequentially transferred to the SM lepton sector. The other scenario is a hierarchical setup with a coupling hierarchy given by $|\lambda_2|\gg |y_2|> |\lambda_1|\sim |y_1|$ and $M_{N_2}\gtrsim M_{N_1}$ (or $|\lambda_3|\gg |\lambda_2| \gtrsim |\lambda_1|$ and $M_{N_3}\gtrsim M_{N_2} \gtrsim M_{N_1}$ for three flavours of right-handed neutrinos). 

The asymmetry generation of the visible and dark sectors are directly related to the size of the physical quantities $\Gamma_{N_1} \epsilon_L$ and $\Gamma_{N_1} \epsilon_{\chi}$ where $\epsilon_i$ is the asymmetry parameter given by
$\epsilon_{L} = \text{Br}(N_1 \rightarrow L \,H) - \text{Br}(N_1 \to \bar{L} \,H^\dagger)$ and
$\epsilon_{\chi} = \text{Br}(N_1 \rightarrow \chi \,\phi) - \text{Br}(N_1 \to \bar{\chi} \,\phi^*)$.
Since the masses and dark couplings are not directly linked, the dark couplings are relatively unconstrained. A small $|\lambda_1|$ coupling allows for large departure of $N_1$ from thermal equilibrium and a large $|\lambda_2|$ coupling allows for large CP violation in the $N_1$ decays since $\Gamma_{N_1} \epsilon_L$ is related to $|\lambda_2|$ and $\Gamma_{N_1} \epsilon_{\chi}$ is related to $|\lambda_2|^2$. The strength of the $|\lambda_2|$ coupling has no effect on the departure from equilibrium of $N_1$ and the $N_2$ decay will have too little departure from thermal equilibrium so the strength of $\Gamma_{N_2} \epsilon_{L,2}$ and $\Gamma_{N_2} \epsilon_{\chi,2}$ are irrelevant.
As such, this hierarchy can allow for sufficient lepton asymmetry generation at lower energy scales than the minimal leptogenesis scenario \cite{Davidson:2002qv}. A mild mass splitting (semi-degenerate RHNs) reduces loop suppressions in the asymmetries allowing for the lowest energy scale. 

\begin{figure*}[t]
\centering
\includegraphics[width=0.8 \textwidth]{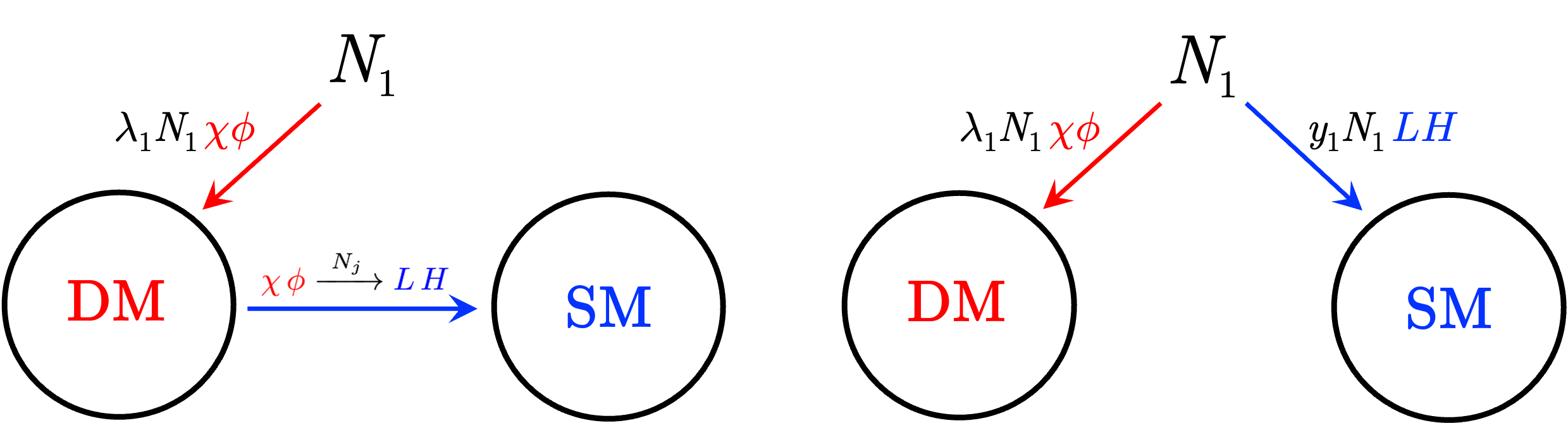}
\caption{Schematic illustration of the two asymmetry-generation mechanisms studied in this work: (left) the wash-in scenario, where an asymmetry first produced in the dark sector is transferred to the Standard Model via scattering, and (right) the co-genesis scenario, where visible and dark asymmetries are generated simultaneously in heavy-neutrino decays, highlighting the dynamical origin of the baryon–dark matter abundance relation.
}
\label{fig:cowash}
\end{figure*}
In Fig.~\ref{fig:cowash} the general framework for the two scenarios is sketched out. These scenarios are unexplored mechanisms of the setup in Ref.~\cite{Falkowski:2011xh}. Since the dynamics of the Boltzmann equations are determined from $\Gamma_i ,$ and $\sigma_{ij}$, decay widths and cross sections (the UV complete forms given in the appendix), we use these 'physical' quantities whenever possible to keep the discussion model independent. 
We first consider a wash-in mechanism, where the asymmetry of the visible sector is generated purely from washout of the asymmetry in the dark sector when $\epsilon_L=0$ such that the only decay channels to the visible sector are CP symmetric. This setup can produce the observed asymmetry provided the Davidson-Ibarra (DI) bound \cite{Davidson:2002qv} is satisfied. 
We then consider a co-genesis scenario similar to Ref.~\cite{Falkowski:2011xh} in which it is shown that a lower bound on the RHN of $10^9$ GeV is needed, matching the Davidson-Ibarra (DI) bound \cite{Davidson:2002qv}. If a large hierarchy is present in the dark couplings with a small hierarchy in the RHN masses, such that all asymmetry is generated from just $N_1$ decays, then a substantially lower bound on the RHN mass is possible resulting in TeV scale leptogenesis. This case was not explored in Ref.~\cite{Falkowski:2011xh}.

It is assumed that no asymmetry is carried in $\phi$ due to washout via fast $\phi \leftrightarrow \phi^\dagger$ interactions. For $\chi$ to be an asymmetric dark matter candidate, the symmetric part  must have a mechanism for it to annihilate away. The thermally averaged annihilation cross section must exceed  $\langle \sigma_{ann} v \rangle \gg 3\times10^{-26} \,\,\text{cm}^3 \,\text{s}^{-1}$\cite{Falkowski:2011xh, Steigman:2012nb} for the symmetric part to be subdominant to asymmetric part.
\\
The minimal setup for this fermionic dark matter scenario with a sufficiently large scalar portal interaction $ \lambda_\phi $ in eq.~(\ref{eq:lag}) to account for the total annihilation of the symmetric part is experimentally ruled out. Alternatively, a scalar dark matter candidate could efficiently annihilate its symmetric component via this mechanism; this could be achieved if the $\phi$ scalar field is lighter than the fermion $\chi$. However, we leave a detailed analysis of this case to future work, and assume that the symmetric dark matter component is efficiently annihilated via another mechanism, such as a dark photon coupling, similarly to Ref.~\cite{Falkowski:2011xh}.
\\
In order to demonstrate how the two leptogenesis mechanisms work, we focus on the simple case with only two RHN flavour states by applying analytical approximations to the Boltzmann equations (the three RHN flavour case discussed later).
\begin{widetext}
\noindent Below, benchmark points are provided that can produce the observed asymmetry which demonstrate certain behaviour for the two mechanisms outlined in this work. 
\begin{table}[h!]
\centering
\caption{Benchmark points for the wash-in scenario. The first benchmark BM1 analytically produces the observed asymmetry (assuming maximum efficiency) and BM2 numerically produces the observed asymmetry.}
{\small
\setlength{\tabcolsep}{5pt}       
\renewcommand{\arraystretch}{1.05} 
\begin{tabular}{|l|c|c|c|c|c|c|c|c|}
\hline
BM & \(M_{N_1}\)[GeV] & \(M_{N_2}\)[GeV] & \(m_{\phi}/M_{N_1}\) & \(m_{\chi}\)[GeV] & \(|\lambda_1|\) & \(|\lambda_2|\) & \(\sqrt{(y^\dagger y)_{11}}\) & \(\sqrt{(y^\dagger y)_{22}}\) \\[0pt]
\hline
BM1 & \(1.6 \times 10^9\) & \(6.1 \times 10^{9}\) & 0.6 & \(7.2 \times 10^{-3}\) & \(1.1 \times 10^{-4}\) & \(\sqrt{4\pi}\) & \(0\) & \(5.1 \times 10^{-3}\) \\[0pt]
\hline
BM2 & \(3.0 \times 10^8\) & \(1.1 \times 10^9\) & 0.6 & \(6.0 \times 10^{-2}\) & \(4.7 \times 10^{-5}\) & \(\sqrt{4\pi}\) & \(0\) & \(2.2 \times 10^{-3}\) \\[0pt]
\hline
\end{tabular}
}
\label{tab:washin}
\end{table}
\begin{table}[h!]
\centering
\caption{Benchmark points for the co-genesis scenario.}
{\small 
\setlength{\tabcolsep}{5pt}        
\renewcommand{\arraystretch}{1.05} 
\begin{tabular}{|l|c|c|c|c|c|c|c|c|}
\hline
BM & \(M_{N_1}\)[GeV] & \(M_{N_2}\)[GeV] & \(m_{\phi}/M_{N_1}\) & \(m_{\chi}\)[GeV] & \(|\lambda_1|\) & \(|\lambda_2|\) & \(\sqrt{(y^\dagger y)_{11}}\) & \(\sqrt{(y^\dagger y)_{22}}\) \\[0pt]
\hline
BM3 & \(6.7 \times 10^6\) & \(9.0 \times 10^6\) & \(\simeq 0\) & 78 & \(1.8 \times 10^{-6}\) & \(2.2 \times 10^{-2}\) & \(1.8 \times 10^{-6}\) & \(2.0 \times 10^{-4}\) \\[0pt]
\hline
BM4 & \(6.7 \times 10^6\) & \(9.0 \times 10^6\) & 0.2 & 0.11 & \(1.8 \times 10^{-6}\) & \(2.2 \times 10^{-2}\) & \(1.8 \times 10^{-6}\) & \(2.0 \times 10^{-4}\) \\[0pt]
\hline
BM5 & 1700 & 2300 & 0.68 & 46 & \(6.2 \times 10^{-8}\) & \(\sqrt{4\pi}\) & \(2.8 \times 10^{-8}\) & \(3.2 \times 10^{-6}\) \\[0pt]
\hline
\end{tabular}
}
\label{tab:cogenesis}
\end{table}
\end{widetext}

\subsection{Boltzmann equations}
The Boltzmann equations describe RHN departure from thermal equilibrium and the generation of the lepton and dark asymmetries in terms of the yield $Y_i=n_i/s(T)$, where $s(T)$ is entropy density and $n_i$ is the number density.
Neglecting $2 \leftrightarrow 2$ scatterings, the Boltzmann equation for RHN departure simplifies to
\begin{equation}
\label{eq:deptboltz}
    \frac{dY_{N_i}}{dz_1}=-\frac{\Gamma_{N_i}}{H_1}z_1\frac{K_1(z_i)}{K_2(z_i)}(Y_{N_i}-Y_{N_i}^{eq}) ,
\end{equation}
where $\Gamma_{N_i}$ is the total decay width (UV expressions in the Appendix) of particle  $N_i$, 
$
    z_i={M_{N_i}}{/T}
$ where $M_{N_i}$ is the mass of the RHN with flavour index $i$, $T$ is temperature,  $H_i\equiv H(M_{N_i})$ is the Hubble rate at temperature $T=M_{N_1}$, $K_1(z_i)$ and $K_2(z_i)$ are modified Bessel functions of the second kind of order $1$ and $2$, and $Y_{N_i}^{eq}$ is the equilibrium yield of particle $N_i$. 
The Hubble rate is $H(T) = \sqrt{8 \pi^3 g_\star /90} \;\, T^2/M_{Pl}$ \cite{Bauer:2017qwy} where $g_*$ is the number of relativistic degrees of freedom in the Standard Model \cite{Husdal:2016haj}.

The Boltzmann equations for lepton and dark matter asymmetries, are:
\begin{align}
 \frac{dY_{\Delta \chi}}{dz_1}
  &= \frac{\Gamma_{N_1}}{H_1}\,\epsilon_{\chi}\,
    \frac{z_1\,K_1(z_1)}{K_2(z_1)}\bigl(Y_{N_1}-Y_{N_1}^{\mathrm{eq}}\bigr)
  \nonumber \\
& \,\,\,
  - \frac{\Gamma_{N_2}}{H_1}\Bigl(2\,\mathrm{Br}_{\chi\,2}^2\,I_{W,\,2}(z_2)\,\frac{Y_{\Delta \chi}}{g_\chi}\Bigr),
\label{eq:dcbltz} &&
\end{align}
\begin{flalign}
    \frac{dY_{\Delta L}}{dz_1} &= 
    \,\frac{\Gamma_{N_1}}{H_1}\,\epsilon_{L} \,\frac{z_1\,K_1(z_1)}{K_2(z_1)}(Y_{N_1}-Y_{N_1}^{eq})\Biggr.  
   \nonumber  \\&- \frac{\Gamma_{N_2}}{H_1}\text{Br}_{L2}\text{Br}_{\chi2}   
    \ {I_{T_{+,\,2}}(z_2)}\,\left(\frac{Y_{\Delta L}}{g_L}+\frac{Y_{\Delta \chi}}{g_\chi}\right)
    \nonumber \\&- \frac{\Gamma_{N_2}}{H_1}\text{Br}_{L2}\text{Br}_{\chi2}  
    \ {I_{T_{-,\,2}}(z_2)}\,\left(\frac{Y_{\Delta L}}{g_L}-\frac{Y_{\Delta \chi}}{g_\chi}\right),
     \label{eq:dlbltz} &&
\end{flalign}
with sub-dominant washout neglected, where the internal degrees of the Lepton doublet $g_L=2$, of the internal degrees of freedom of the dark fermion singlet $g_\chi=1$, of the Majorana RHN $g_N=2$, $\epsilon_L$ and $\epsilon_\chi$ are the asymmetry parameters, 
$\text{Br}_{a\,i}$ are $N_i$ branching ratios, and $I_{(W,T\pm),\,2\,}(z_2)$ are the $2 \leftrightarrow 2$ washout and transfer scattering terms mediated by $N_2$. The washout and scattering terms are given by
\begin{equation}
\label{eq:thermcross}
    I(z_i) ={z_i^3}\frac{z_1}{z_i}\int_{\hat{s}_{min}}^{\infty}d\hat{s}\frac{\sigma_{ab}(\hat{s})p_{cm}^2(\hat{s}) \sqrt{\hat{s}}\,{K}_1\left( z_i\,\sqrt{\hat{s}}  \right)}{16 \pi^2\hat{\Gamma}_i\text{Br}_a \text{Br}_b},
\end{equation}
\\
where $\hat{s}_{min}=(r_x+r_y)^2$, $\hat{\Gamma}_i=\Gamma_{N_i}/M_{N_i}$, $\sigma_{ab}$ is the cross section, $a,b\in\{L,\chi\}$, $p_{cm}$ is the centre of mass momentum,
$
    r_x  = {m_x}{/M_{N_i}}
$
and 
$
    \hat{s}={s}{/M_{N_i}^2}
$ similar to \cite{Falkowski:2017uya}
where $s$ is the Mandelstam variable centre of mass Energy (see Appendix for the full cross sections). 

\subsection{Analytical Approximations \label{sec:analytic}}
In this section we provide analytical approximations to demonstrate how the RHN mass bounds in the wash-in and co-genesis regimes arise, and build an intuition for the existing scale relations. 
\subsubsection*{Out of equilibrium analytical solution}
The out of equilibrium solution for $N_i$ (RHNs) drive asymmetry generation for $L$ and $\chi$.
While there is no single analytic approximation that can describe $N_i$ departure from equilibrium as a continuous function over all $z_i$, it is possible to solve for $Y_{N_i}$ analytically across different regions.
The equilibrium yield is approximated using analytical approximations for early and late time behaviour as 2 regions ($z_i<z_0$ and $z_i > z_0$). This split is chosen at $z_0 = 1.5$ since the large $z_i$ approximation, from the first order terms of the large $_i$  expansion, is invalid for $z_i<z_0$. From the following:
\begin{equation}
\label{eq:YNeqap}
Y_{N_i}^{eq}=
    \begin{cases}
        \scalebox{0.95}[1]{$\tfrac{45}{2\pi^{4}g_{*s}} 
        \left[2-\left(2 z_0^{-2}+K_2(z_0)\right) z_i^2\right]$}, & z_i<z_0\\
        A\,e^{-z_i} z_i^{3/2}, & z_0<z_i,
    \end{cases}
\end{equation}
the Boltzmann equations in eq.~(\ref{eq:deptboltz}) can be analytically solved, 
where $A = A(z_0=1.5) \simeq 6.93\times10^{-3}$ is constant for a fixed value of $z_0$ (see eq.~(\ref{eq:approxconst}) in section
\ref{sec:apdef} of the Appendix
for the full functional dependence) and $g_{\star s}$ is the effective number of relativistic degrees of freedom for entropy. 

To solve eq.~(\ref{eq:deptboltz}) analytically, the departure is assumed to be a first order perturbation with a small coefficient such that higher order terms are suppressed.
In the following, for small $\hat{\Gamma}_i =\Gamma_{N_i}/M_{N_i}$, the solution for the yield can be expanded as
\begin{equation}
    Y_{N_i} \simeq Y_{N_i}^{\,eq} +\frac{1}{\hat{\lambda}_i}\,Y_i,
\end{equation} 
where $\hat{\lambda}_i= M_{N_i}/H_1$.
Using eq.~(\ref{eq:YNeqap}), an analytical approximation for $Y_{N_i}$ is derived by
dividing the solution into different regions depending on the dominant terms in eq.~(\ref{eq:deptboltz}) for different regions of $z_i$.

These dominant terms correspond to different physical behaviour:
when $Y_{N_i}$ is in thermal equilibrium (Region 0), is in freeze-out (Region I), after freeze-out ends and it starts to return to thermal equilibrium (Region II), and after it has returned to thermal equilibrium (Region III). 
The point at which freeze-out ends, $z_f$, can be approximated for small $\Gamma_{N_i}/H_i$ as 
$
    z_f\simeq z_0+\hat{\lambda}_i\hat{\Gamma}_{i}{9}{/4}
$

and the point which it recombines at, $z_{r}$, is approximately
$
    z_r \simeq {2}{/(\hat{\lambda}_i\hat{\Gamma}_{i})}
$. 
The analytical solution for the departure from thermal equilibrium is then given by:
\begin{widetext}
\noindent
\begin{equation}
\label{eq:YNan}
Y_{N_i}\simeq
\begin{cases}
    Y_{N_i}^{eq}, & {\color{pale-orange}\text{Region 0}}:\;z_i<1.5\\
    A e^{-z_i} \left[z_i^{\frac{3}{2}} + \frac{z_i^{-1/2}}{\hat{\lambda}_i\hat{\Gamma}_{i}}(z_i-3/2) \right], &  {\color{red}\text{Region I}}\,:\; 1.5<z_i<z_f \\
    A e^{-z_i} \left[z_i^{\frac{3}{2}} + \frac{z_f^{-1/2}}{\hat{\lambda}_i\hat{\Gamma}_{i}}e^{\left( {\hat{\lambda}_i\hat{\Gamma}_{i}} \,(z_f^2-z_i^2)/2 -  (z_f-z_i) \right)} \left( z_f-3/2\right)\right], & {\color{gray}\text{Region II}}:z_f\,<\,z_i<\,z_{r} \\
    Y_{N_i}^{eq}, & {\color{light-purple}\text{Region III}}:z_r<z.
\end{cases}
\end{equation}
\end{widetext}

\begin{figure}[!htbp]
\centering
\includegraphics[width=0.47\textwidth]{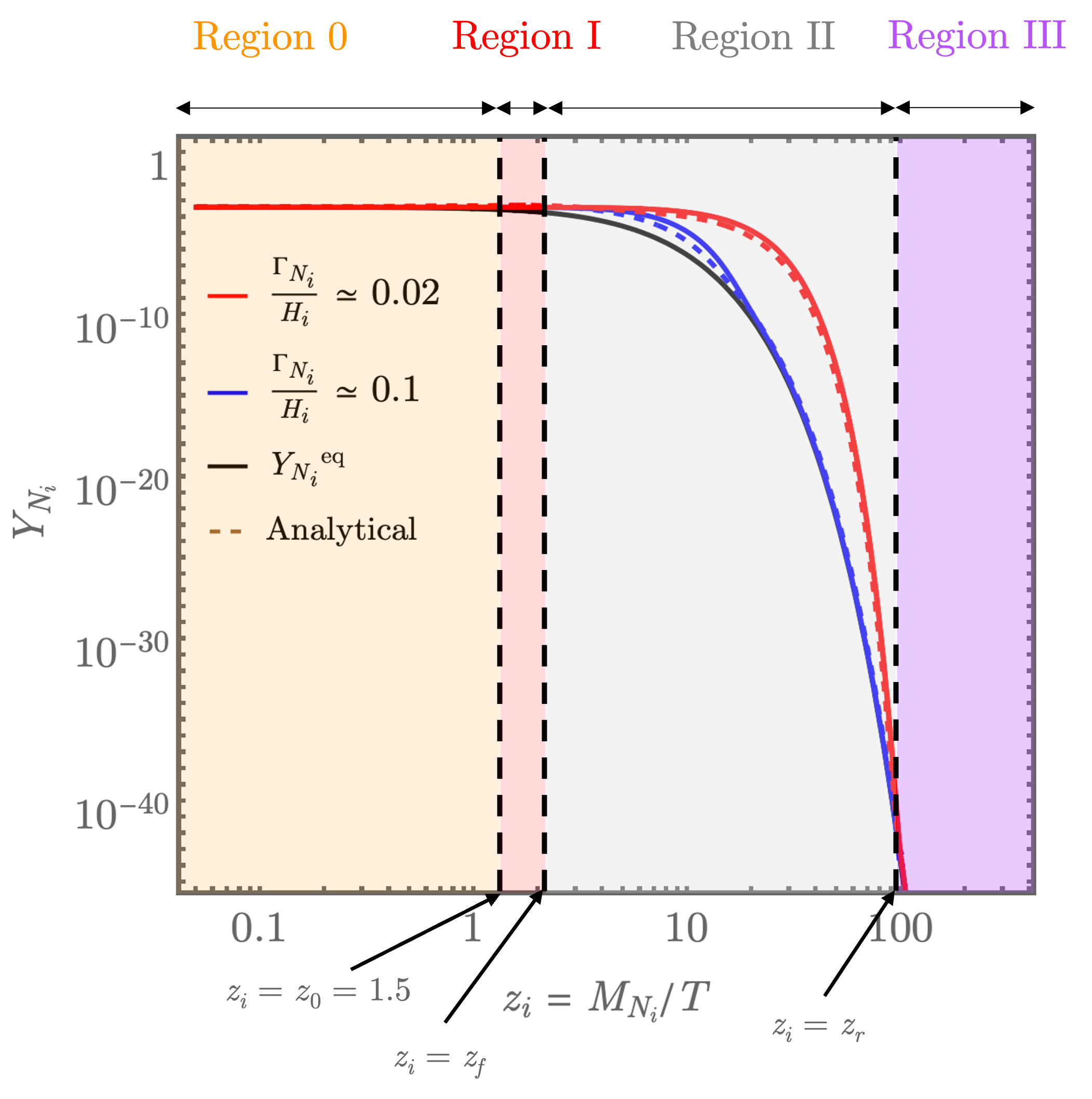}
\caption{
Departure of the heavy-neutrino abundance from equilibrium as a function of $z=M_{N_i}/T$, showing numerical (solid) and analytical (dashed) solutions for different $\Gamma_{N_i}/H_i$, where smaller decay rates produce larger departures from equilibrium and therefore more efficient asymmetry generation.}
\label{fig:Regions}
\end{figure}
In Fig.~\ref{fig:Regions} we demonstrate that $\Gamma_{N_i}/H_i\propto 1/(Y_{N_i}-Y_{N_i}^{eq})$. This approximation, which is only valid for ${\Gamma_{N_i}}{/H_i}\lesssim 0.16$, matches closely with the numerical solution for eq.~\ref{eq:deptboltz}. The red line in Fig.~\ref{fig:Regions} shows the $N_1$ yield for BM1-BM4. The wash-in and co-genesis mechanisms, outlined below, arise following the case $\Gamma_{N_1}/M_{N_1}\ll \Gamma_{N_2}/M_{N_2}$ allowing for larger departure and larger CP mixing. As seen in Fig.~\ref{fig:Regions}, a smaller $\Gamma_{N_i}$ results in more departure from equilibrium characterised by $Y_{N_i}-Y_{N_i}^{eq}$ in eqs.~(\ref{eq:dcbltz},\ref{eq:dlbltz}) with $Y_{N_i}^{eq}$ given in the appendix in eq.~(\ref{eq:YNeq}). This case provides significantly more asymmetry generation from $N_1$ decays than from both $N_1$ and $N_2$ decays with similar strengths. 
\\
With this out of equilibrium analytical solution to eq.~(\ref{eq:deptboltz}), approximate solutions for the lepton and dark matter asymmetries are then derived in the next section for negligible washout.
\subsubsection*{Asymmetry generated neglecting washout}

In the following, we provide analytical approximations for the generation of the lepton and dark asymmetries without washout present in order to derive the mass bounds.
Initially, the washout and asymmetry generation are in equilibrium, then the washout becomes subdominant to the asymmetry generation due to the inverse decays becoming kinematically forbidden. The behaviour in this asymmetry generating region can be approximated since washout is negligible. 

Using $Y_{N_i}$ in Region II from eq.~(\ref{eq:YNan}), the following formula for the asymmetries of $L$ and $\chi$ in the region when no washout is present (i.e. survival fraction of $\eta=1$) can be derived, valid for $z_1>z_f$ 
\begin{equation}
\label{eq:anasym}
    Y_{\Delta a}^{\eta=1}(z_1)\approx {\epsilon_a \,Y_{N_1}^{eq}(0)}\cdot \left(1-e^{(z_f^2-z_1^2)\hat{\lambda}_1\hat{\Gamma}_1/2}   \right),
\end{equation}
where $a=L, \chi$. If the washout processes remains subdominant to the asymmetry generation, then this is a valid approximation for the asymmetry generation. However, for the washout present in this model, this is not the case. The critical point, when washout becomes dominant over the asymmetry generation, is $z_c$. Smaller $z_c$ indicate more washout domination. The effect of washout on the final asymmetries can be found from asymptotic solutions to the Boltzmann equations at $z_c$.
In the next section, we provide the analytical behaviour of the $2 \leftrightarrow 2$ scattering processes when asymmetry generation becomes sub-dominant to washout.
\subsubsection*{Analytical Scattering}
We present the analytical approximations for the dominant washout processes to later derive the asymptotic solution for the asymmetries.  
The $2 \leftrightarrow 2$ scattering washout and transfer processes have analytical solutions from \cite{Falkowski:2011xh} split between early $z_i$ and late $z_i$.
The $2 \leftrightarrow 2$ scattering functions in the early $z_i$ regime have the same behaviour when normalised by the Branching ratios due to the Breit–Wigner width (due to all having the same resonance at $s=M_{N_i}$).
In the small $z_i$ regime, from Ref.~\cite{Falkowski:2011xh}
\begin{equation}
\label{eq:BWwidth}
    I_{\,W,\,T_\pm}(z_i)\approx\frac{z_i^3}{4}K_1(z_i),
\end{equation}
and for the late $z_i$ regime; $I_{W_1,\,W_2,\,T_+}$ is derived from integrating Taylor series expansions about $s=0$, $I_{W_1}$ is from washout with only one scalar in the initial state and $I_{W_2}$ is with 2 scalars in the initial or final state, $I_{T+}$ is the mixed sector lepton-violating transfer. Here, $m_\phi$ and $m_\chi$ are assumed to be negligible. The lepton-conserving transfer $I_{T-}$ is negligible for late $z_i$: 
\begin{equation}
\begin{gathered}
I_{(W_{1},T_{+})_i}(z_i)\approx 
    \frac{32}{\pi}\frac{\Gamma_{N_i}}{M_{N_i}}\frac{z_1}{z_i^3},  I_{W_{2,\,i}}(z_i)\simeq
    \frac{16}{\pi}\frac{\Gamma_{N_i}}{M_{N_i}}\frac{z_1}{z_i^3}. 
\end{gathered}
\end{equation}
Only the dominant contributions to the washout are given here, the sub-dominant $T_-$ analytic solution is provided in the appendix.
In the next section, we demonstrate the limits on the wash-in mechanism using these analytical solutions.
\subsection{Wash-in Scenario}
In the limit $\epsilon_L\rightarrow0$ (with hierarchy $|\lambda_2|\gg|\lambda_1|$ and $M_{N_2}\gtrsim M_{N_1}$) a lepton asymmetry can be produced from transfer of the dark asymmetry to the lepton sector via $2 \leftrightarrow 2$ scatterings.
We first consider this wash-in scenario in the limit $m_\phi=0$ and then, at the end of this section, we discuss the effect of $m_\phi \lesssim M_{N_1}$.
The washout at late $z_1$ is dominated by $I_{W_\chi,2}$ for $Y_{\Delta \chi}$ and $I_{T_+,2}$ for $Y_{\Delta L}$.
For $z_1>z_c$, from Ref.~\cite{Falkowski:2011xh}, these dominant terms yield the asymptotic lepton asymmetry as 
\begin{equation}
\begin{split}
    Y_{\Delta L}^\infty &\gtrsim -\frac{\text{Br}_{L2}}{3}\,(e^{-\text{Br}_{L2} \frac{3}{4}\frac{W_2}{z_c}}-e^{-3\frac{W_2}{z_c} })\,Y_{\Delta \chi}^{\eta_\chi =1}(z_c) \\
    & > -\frac{\text{Br}_{L2}}{3}\,(\eta_L=1)\,Y_{\Delta \chi}^{\eta_\chi =1}(z_c),
\end{split}
\end{equation}
where $W_i=({32\, \hat{\Gamma}_i}{/\pi})({\Gamma_{N_i}}{/H_1})(M_{N_1}/M_{N_i})^3$, $\text{Br}_{\chi\,2}\simeq1$,  and the asymptotic dark matter asymmetry is then given by 
\begin{equation}
    Y_{\Delta \chi}^{\infty}\simeq e^{-3\,W_2/z_c}\,Y_{\Delta \chi}^{\eta =1}(z_c),
\end{equation}
where only $W_2$ is present since subdominant terms are neglected. 
Using $Y_{\Delta \chi}(z_c)$ from eq.~(\ref{eq:anasym}), we obtain a lower bound on $M_{N_1}$ for sufficient asymmetry from wash-in by assuming a maximum production factor of $\eta_L=1$.

Similarly to the Davidson-Ibarra bound \cite{Davidson:2002qv} this wash-in bound depends on the UV completion of the model, and to generate the observed asymmetry the bound
\begin{equation}
    M_{N_1}\gtrsim \frac{24 \,\pi\,v_{ew}^2}{m_3}\cdot \frac{Y_{\Delta L,\text{obs}}^{\infty}}{Y_{N_1}^{eq}(0)}\cdot \frac{1}{\eta_L} \approx \frac{10^9}{\eta_L}\, \text{GeV},
\end{equation}
must be satisfied, where $\eta_L$ is the wash-in production fraction of DM to leptons, $v_{ew}$ is the vev given as $v_{ew}\simeq174$ GeV, normal hierarchy is assumed with type I seesaw. If $m_\phi/M_{N_1}\lesssim1$, a wash-in production fraction of $\eta_L\sim1$ is possible due to more dark asymmetry generation. Therefore, the bound on the RHN mass is such that $M_{N_1}\gtrsim 10^9\, \text{GeV}$ with a possible minimal mass model shown in BM1. 
Numerically, the bound is weaker than the analytical estimate due to non-negligible $m_\phi$ effects (see BM2), as discussed later.

The mass of the dark matter in this scenario needed is:
\begin{equation}
    m_\chi = \frac{5}{2}\frac{Y_{\Delta L}^{\infty}}{Y_{N_1}^{eq}(0)}\frac{M_{N_2}}{\Gamma_{\chi,\,2}}\frac{2+\kappa}{\kappa}\frac{M_{N_2}}{M_{N_1}}\frac{e^{\hat{\Gamma}_1\hat{\lambda}_1\,9/4}}{(1-e^{-z_r})}\frac{(\text{GeV})}{\eta_\chi},
\end{equation}
where $Y_{\Delta L}^{\infty}$ is the expected lepton asymmetry, $\eta_\chi$ is the survival fraction from washout, $\kappa=\text{Br}_{\chi\,1}/\text{Br}_{L\,1}$. The optimal value for $\kappa\approx1$, this provides a reasonable balance between asymmetry generation and departure from thermal equilibrium.
Assuming optimal parameters are valid, the minimum value of the mass of $\chi$ is found to be $m_\chi \gtrsim 5\times 10^{-6}$ GeV.
For DM stability, the mass of $\chi$ has to be lower than $M_{N_1}$ or $m_\phi$ so the maximum $\chi$ mass is bounded by $m_\chi \lesssim \text{min}[M_{N_1},m_\phi]$.
Thus, the DM mass range in this scenario is:
\begin{equation}
    \frac{5\times10^{-6}}{\eta_\chi}(\text{GeV})\lesssim m_\chi \lesssim \text{min}[M_{N_1},m_\phi].
\end{equation}

We observe that the value of the RHN mass bound obtained here roughly coincides with the DI bound despite stemming from a different mechanism. The next section explores a mechanism that leads to a much lower right-handed neutrino mass bound, the co-genesis scenario.

\subsection{Co-genesis scenario}
In this co-genesis regime, CP violation in both channels occur, $\epsilon_L, \epsilon_\chi \neq 0$ (with hierarchy $|\lambda_2|\gg|\lambda_1|$ and $M_{N_2}\gtrsim M_{N_1}$). $N_1$ decays dominate asymmetry generation to both sectors and $N_2$ interactions dominate washout.
Assuming weak washout in the lepton sector, the asymptotic lepton asymmetry is:
\begin{equation}
    Y_{\Delta L}^\infty \approx e^{-(3/4)\cdot\text{Br}_{L\,2}\,W_2/z_c}\,Y_{\Delta L}^{\eta =1}(z_c),
\end{equation}
where $\text{Br}_{\chi\,2}\simeq1$, and $z_c$ is the critical point where washout dominates lepton asymmetry generation.

Since the RHN mass bound in this scenario depends on the size of the mass splitting between the RHN states, we define $ r_{21}=M_{N_2}/M_{N_1} $.
Avoiding the resonant leptogenesis regime ($ M_{N_2}-M_{N_1} > \Gamma_{N_2}$) \cite{Pilaftsis:2003gt}, we use the smallest $r_{21}\simeq 1.35$ that is valid for couplings up to the perturbative maximum. Assuming a survival fraction of $\eta=1$,
the bound is
\begin{equation}
\label{eq:anbound}
    M_{N_1}\gtrsim r_{21}\,\frac{144\,\pi\, v_{ew}^2}{m_3}\left(  \frac{Y_{\Delta L}^{\infty}}{Y_{N_1}^{eq}(0)}  \right)^2 \simeq 1 \,\,\text{TeV}.
\end{equation}

However, this approximation does not reflect what the true mass bound is when taking into account washout and more importantly DM stability, but it demonstrates that a lower mass bound, around the TeV scale, is feasible. To find the true physical bound, the point where the $\chi$ mass is maximal while the $N_1$ mass is minimal, needs to be found.

Due to the strong hierarchy in this regime, the DM has very strong washout, to the point where the washout dictates the DM asymmetry for minimal $N_1$ mass. As such the initial asymmetry generated cannot be assumed to be given by $Y_{\Delta \chi}^{\eta=1}=\epsilon_\chi \,Y_{N_1}(z_1=0) $, instead the DM abundance can be approximately solved as a Saha equation to find the critical point $z_c$ where the asymmetry generation becomes subdominant to the washout by solving for $Y_{\Delta \chi}$ when $dY_{\Delta \chi}/dz_1=0$ in eq.~(\ref{eq:dcbltz}). 
\newline
\indent The initial abundance of DM generated is assumed to be given at the point where the Saha decouples from equilibrium. Though there may be more asymmetry generation after $z_s$, this behaviour cannot be analytically described so some asymmetry generation is not accounted for. The Saha critical point for DM ($z_s$) is determined by solving $\Gamma_{N_2}/H_1\, I_{W,\,2}(z_s\cdot r_{21})=1$ with $z_2= M_{N_2}/M_{N_1}\cdot z_s$, which is the point where the washout becomes subdominant to Saha equation. For the initial DM abundance, the Saha solution given by
\begin{equation}
\label{eq:saha}
    Y_{\Delta \chi}(z_s)\simeq \frac{\Gamma_{N_1}}{\Gamma_{N_2}}\frac{\epsilon_\chi z_s}{2} \frac{K_1(z_s)}{K_2(z_s)} \frac{(Y_{N_1}(z_s)-Y_{N_1}^{eq}(z_s))}{\text{Br}_{\chi\,2}^{\,\,\,2}\, I_{W,\,2}(z_{2,\,s})} \,. 
\end{equation}

From this, the critical point $z_c$, for DM at the point  where the asymmetry generating term becomes subdominant to the $2 \leftrightarrow 2$ washout term, is found assuming that $Y_{\Delta \chi}(z_c)=Y_{\Delta \chi}(z_s)$ in eq.~(\ref{eq:dcbltz}).
The asymptotic asymmetry is given as 
\begin{equation}
\label{eq:asympchi}
    Y_{\Delta \chi}^{\infty}(z_\infty)\simeq e^{-{3\,W_2/}{z_c}}\,Y_{\Delta \chi}(z_c),
\end{equation}
where $\text{Br}_{\chi\,2}\simeq 1$. If $z_s>z_c$, such that the washout is so strong the asymmetry generating term and the washout are never out of equilibrium, then the Saha solution dictates the final asymmetry. 
Looking at BM3, a point close to the bound, using the analytical calculation produces a lepton asymmetry of $Y_{\Delta L}/Y_{\Delta L}^\infty \simeq 1.6$ and gives $m_\chi\simeq$ 2.7 TeV. 
\\
If the masses of $\chi$ and $\phi$ are assumed to be negligible, then the bound on $N_1$ was found to be $M_{N_1} \gtrsim 10^7$ GeV for stable DM, well above the possible bound of $M_{N_1} \gtrsim 1$ TeV.
This is due to the extreme washout present in the DM sector, which destroys generated DM asymmetry generated disallowing stable DM below $\sim 10^7$ GeV since the DM asymmetry is inversely proportional to the DM mass. From eq.~(\ref{eq:anbound}), ignoring DM stability would allow for a significantly lower bound. To reach this bound, the dark washout would need to be suppressed to generate a larger DM abundance. Next we will demonstrate that if the $\phi$ mass is non-negligible, it can kinematically suppress the dark washout.
\\
\subsection*{Washout Suppression from non-zero \texorpdfstring{$\phi$}{phi} mass} 
We provide the analytical behaviour for the $2 \leftrightarrow2$ scattering terms with the suppression effect from $m_\phi$ (assuming $m_\chi \ll M_{N_1}$). The small $z_i$ regime is the same as in eq.~(\ref{eq:BWwidth}),
and for the late $z_i$ regime; $I_{T_+,\,i}$ is derived from integrating Taylor series expansions about $\hat{s}=r_{\phi\,i}=m_\phi/M_{N_i}$ , $I_{W_1,\,i}$ is approximated and set $I_{W_2,\,i}=0$ since it is negligible compared to $I_{W_1,\,i}$ for $r_{\phi\, i}\gtrsim 0.1$ ($I_{T_-,\,i}$ is provided in the appendix but is negligible here). The behaviour of the washout for late $z_i$ with $r_{\phi\,i} \gtrsim 0.1$ is
\begin{equation}
\label{eq:washphi}
    I_{W_{1,\,i}}(z_i)\approx\frac{32}{\pi}\frac{\Gamma_{N_i}}{M_{N_i}}\frac{r_{\phi \, i}}{z_i}K_1(z_i \cdot \,r_{\phi \,i}), 
\end{equation}
\begin{equation}
\begin{split}
\label{eq:transphi}
    I_{T_{+,\,i}}(z_i)\simeq\frac{4}{\pi}\frac{\Gamma_{N_i}}{M_{N_i}}\frac{ r_{\phi \, i}}{ z_i} \bigl(&(r_{\phi \, i}^2 z_i^2+8) K_1(r_{\phi \, i} z_i)\\&\,\,\,\,+4 r_{\phi \, i} z_i K_0(r_{\phi \, i}
   z_i)\bigr),
   \end{split}
\end{equation}

where $r_{\phi \, i}=m_{\phi}/M_{N_i}$. In the limit of $r_{\phi \, i}\rightarrow 0$,  analytical formula for the late $z_i$ regime from Ref.~\cite{Falkowski:2011xh} is recovered.
Examining eqs.~(\ref{eq:asympchi}, \ref{eq:washphi}) and (\ref{eq:transphi}), the effect of the $\phi$ mass suppression on the washout can be seen. The kinematic effect from the finite $\phi$ mass shifts the $z_c$ point to later times and lower temperatures resulting in a later onset of $2 \leftrightarrow 2$ washout domination. However, it simultaneously suppresses the decay rate and asymmetry parameters resulting in an intricate interplay of the involved processes. The total decay width of $N_i$ is now given by
\begin{equation}
    \Gamma_{N_{i}} = \frac{2(y_i^{\dagger}y)_{ii} }{8 \pi g_N}M_{N_i} +( 1-r_{\phi \, i}^2)^2\frac{|\lambda_i|^2}{8 \pi g_N}M_{N_i} ,
\end{equation}
where the internal degree of freedom of the RHN $g_N =2$, for the asymmetric decay $N_1\rightarrow LH$ the asymmetry parameter is
\begin{widetext}
\begin{equation}
\label{eq:fullepsL}
    \epsilon_{L,\,1}\simeq\frac{1}{8\pi  }\frac{\text{Br}_{L \,1}}{2(y^{\dagger}y)_{11}}\left( \frac{M_{N_1}}{M_{N_2}}\text{Im}\left[(y^\dagger y)_{12}^{2} \right]+\frac{M_{N_1} M_{N_2}}{M_{N_2}^2-M_{N_1}^2}\text{Im}\left[2(y^\dagger y)_{12}^{2}+(y^\dagger y)_{12}\,\lambda_1^{*} \lambda_2(1-r_{\phi\,1}^{\;2})^2\right]\right),
\end{equation}
and for the asymmetric decay $N_1\rightarrow \chi \phi$ the asymmetry parameter is
\\
\begin{equation}
    \epsilon_{\chi,1}\simeq\frac{1}{16\pi }\frac{\text{Br}_{\chi \,1}}{|\lambda_1|^2}\left( \frac{M_{N_1}}{M_{N_2}}\text{Im}\left[(\lambda_1^* \lambda_2)^2 \right](1-r_{\phi1}^2)^2
    +\frac{M_{N_1} M_{N_2}}{M_{N_2}^2-M_{N_1}^2}\text{Im}\left[(\lambda_1^* \lambda_2)^2(1-r_{\phi1}^2)^2
    +2(y^\dagger y)_{12}\lambda_1^* \lambda_2\right]\right),
\end{equation}
\end{widetext}
where the parametrisation of $(y^\dagger y)$ is given in the appendix. 

In the above equations, the $L,H,\chi$ masses are taken to be negligible, and the analytic description demonstrates that in principle the dark washout can be suppressed given a finite $\phi$ mass. This results in more dark asymmetry generation and a lower RHN mass. However, the asymptotic solution demonstrated above for the washout effect is invalid when $r_{\phi \, i}$ is present, since the integrals of eqs.~(\ref{eq:washphi}) and (\ref{eq:transphi}) have no closed-form expressions. Hence, to study the effect of $r_{\phi \, i}$ and the physical washout, we turn to numerical analysis in the next section.

\subsection*{Thermalisation of \texorpdfstring{$N_1$}{N1} at the low scale.}
At low masses and very small couplings $|\lambda_1|,\,|y_1|$, the interaction rate of $N_1$ may fall below the Hubble expansion rate around $T \simeq M_{N_1}$, making it difficult for $N_1$ to reach its equilibrium abundance in the minimal setup. One simple possibility to restore thermalisation is to extend the model by a singlet scalar field $S$ and include a renormalisable coupling of the form
\begin{equation}
\mathcal{L} \supset \kappa\, N_1 N_2 S + \text{h.c.},
\end{equation}
which induces the two-body decay $N_2 \to N_1 S$. Since $N_2$ is efficiently thermalised through its sizeable couplings $|\lambda_2|, |y_2|$, even a small mixing coupling $\kappa \sim 10^{-7}$ is sufficient for $N_2$ decays to populate $N_1$, effectively bringing $N_1$ close to thermal abundance before freeze-out. In this case, the co-genesis mechanism proceeds as genuine thermal leptogenesis even at the TeV scale. Alternatively, a non-thermal but effectively ``thermal'' initial abundance of $N_1$ can arise if the inflaton couples directly to the right-handed neutrino sector and reheating produces $N_1$ with an abundance comparable to or larger than its equilibrium value. Both possibilities ensure that $N_1$ is present in the early Universe with a population consistent with the assumptions of our Boltzmann analysis, without modifying the dynamics responsible for the generation of the visible and dark asymmetries. Later we will also discuss that the singlet scalar $S$ can play an important role in the annihilation of the symmetric DM component in this scenario.

\section{Numerical analysis}
To accurately demonstrate the effect of $r_{\phi\,1}$ on washout, the Boltzmann equations and the thermally averaged cross sections are solved numerically.

\begin{figure}[!htbp]
\centering
\includegraphics[width=0.45\textwidth]{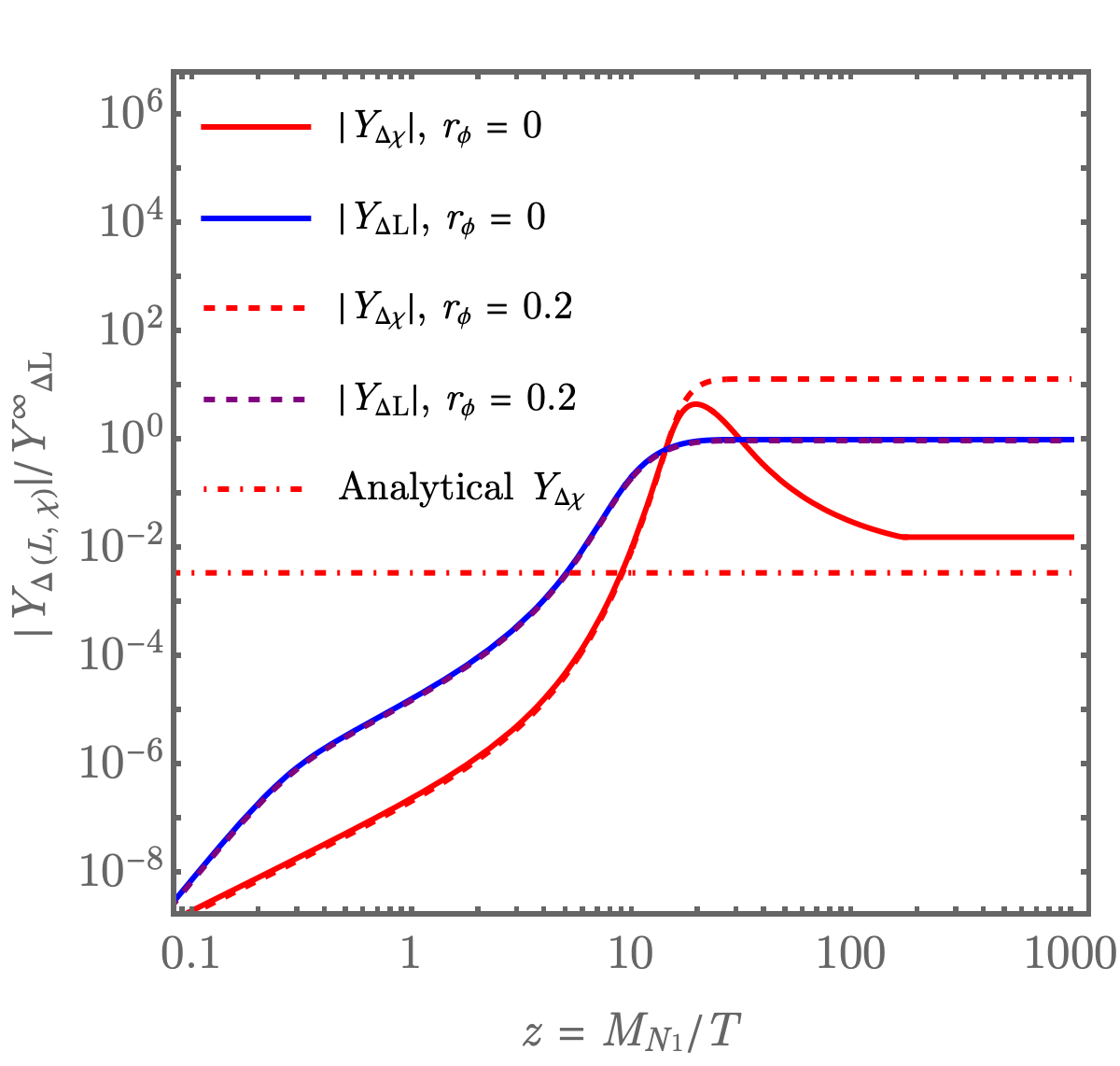}
\caption{Effect of increasing the scalar mass ratio $m_\phi/M_{N_1}$ on dark-sector washout, demonstrating that larger $m_\phi$ suppresses late-time $2\leftrightarrow2$ washout processes,
allowing a larger dark asymmetry to survive and enabling successful leptogenesis at lower RHN mass scales.
The solid lines correspond to BM3 with no dark-washout suppression, where
$m_\phi/M_{N_1}=r_{\phi 1}\simeq 0$ and $m_\phi\gtrsim m_{\chi}\simeq 78~\mathrm{GeV}$.
The dashed lines correspond to BM4 with dark-washout suppression, where
$r_{\phi 1}=0.2$, $m_{\phi}\simeq 1.7\times 10^{6}~\mathrm{GeV}$, and
$m_{\chi}\simeq 0.11~\mathrm{GeV}$.
The red dot-dashed horizontal line shows the asymptotic analytic estimate of BM3 for dark asymmetry from
eqs.~(\ref{eq:saha}) and (\ref{eq:asympchi}).}
\label{fig:mpeff}
\end{figure}

\begin{figure*}[!htbp]
\centering
\includegraphics[width=0.95\textwidth]{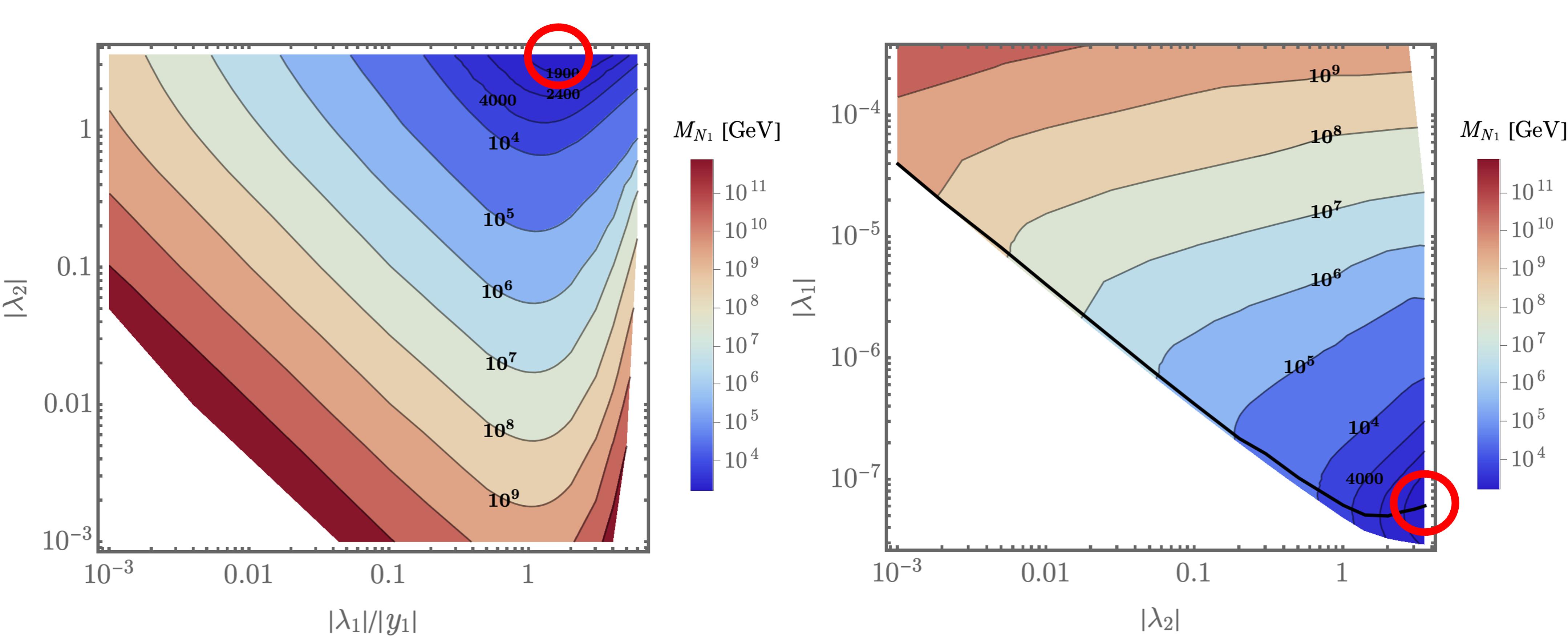} 
\caption{Contours of the minimum RHN mass $M_{N_1}$ required to reproduce the observed asymmetry as a function of coupling ratios, illustrating how hierarchical dark-sector couplings lower the leptogenesis scale, with the highlighted point indicating the parameter region achieving the minimal viable mass. Left demonstrates the effect of the ratio $|\lambda_1|/|y_1|$ and $|\lambda_2|$ on the
$N_1$ mass bound. $|\lambda_1|/|y_1|\lesssim 1$ has less CP violation, $|\lambda_1|/|y_1|\gtrsim1$ has less departure from thermal equilibrium.  Right shows how the bound changes as the hierarchy between $|\lambda_2|$ and $|\lambda_1|$ increases, with the black line showing the minimum $N_1$ mass for the corresponding $\lambda_2$ coupling. The red circle highlights the parameter point where the bound is lowest. 
The mass gap is set to $M_{N_2}/M_{N_1}\simeq 1.35$, $|y_1|\approx\sqrt{M_{N_1}}\cdot 6.8\times 10^{-10}/\sqrt{[\text{GeV}]}$, $|y_2|\approx\sqrt{M_{N_2}}\cdot6.5\times 10^{-8}/\sqrt{[\text{GeV}]}$.}
\label{fig:contour}
\end{figure*}

In Fig.~\ref{fig:mpeff}, an increase in $r_{\phi\,1}$ clearly suppresses the dark washout. The negative effect on the lepton asymmetry is minimal. For $r_{\phi1}\gtrsim0.01$, washout suppression becomes non-negligible. 
For $r_{\phi\,1} =0.2$, significantly more dark asymmetry survives compared to the lepton asymmetry lost.
\\
\\
With a large hierarchy in $|\lambda_1|\ll|\lambda_2|$, without washout suppression, the washout on the DM asymmetry becomes too small for lower $M_{N_1}$ values resulting in a bound of about $10^6$ GeV while still having stable DM as if $|\lambda_2|$ is too large, then the dark washout becomes too large. However, with a large hierarchy in $|\lambda_1|\ll|\lambda_2|$ with washout suppression, the RHN mass bound is able to be significantly lowered as $|\lambda_2|$ can be increased to the perturbative maximum.
As shown in section \ref{sec:analytic}, $|\lambda_1|$ must be small: setting $|\lambda_1|$ near the perturbative maximum would erase the $N_1$ departure from thermal equilibrium.
The optimal choice is $|\lambda_1| \sim |y_1|$; $|\lambda_1|\gg |y_1|$ suppresses departure from thermal equilibrium, while $|\lambda_1|\ll |y_1|$ suppresses CP violation. As the bound on $M_{N_1}$ decreases, $|y_1|$ decreases as $(y^\dagger y)_{11} \propto M_{N_1}$ and the optimal value of $|\lambda_1|$ tracks it.
\\
\\
In Fig.~\ref{fig:contour}, the effect on how the mass bound changes depending on the choice of parameters for the ratio of $|\lambda_1|/|y_1|$ against $|\lambda_2|$ is demonstrated. From Fig.~\ref{fig:contour}, for the majority of the parameter space, the optimal value of $|\lambda_1|/|y_1| \simeq1$ except for RHN masses below ~4 TeV where the optimal value begins to shift to $|\lambda_1|/|y_1| \simeq2.2$ at the perturbative maximum.
As the bound approaches about 4 TeV, since the sphaleron process freezes out near $T\simeq 130$ GeV \cite{DOnofrio:2014rug}, any lepton asymmetry generated after this does not get transferred to the baryon sector. This is partially responsible for the shift for the optimal value to $|\lambda_1|/|y_1|\rightarrow2.2$ when $|\lambda_2|=\sqrt{4 \pi}$ due to slightly more asymmetry generation at lower temperatures since the slightly stronger washout that occurs after EWSB becomes irrelevant.
\\
\\
In Fig.~\ref{fig:contour}, the effect of how the bound changes with the hierarchy of $|\lambda_2|$ against $|\lambda_1|$ is shown. As the mass bound decreases, the $|\lambda_1|$ coupling also decreases. At the RHN mass bound, the ratio of $|\lambda_2|/|\lambda_1|$ is about $\sim \mathcal{O}(10^8)$. 
Though this hierarchy is extreme, it does not violate radiative stability. There are no unsuppressed corrections to $|\lambda_1|$ from $|\lambda_2|$; loops that feed $|\lambda_2|$ into $|\lambda_1|$ always include a $|y_1|$ or $|\lambda_1|$ insertion and are therefore parametrically suppressed, so no radiative instability arises. The red circle highlights an interesting parameter point which provides the lowest mass scale scenario for the RHNs around $\sim \mathcal{O}(2) \, \rm TeV$. The next section shows how the mass of $\chi$ varies with respect to washout suppression in this interesting parameter range.
\subsection*{Dark Matter mass range in TeV scale leptogenesis}
\begin{figure}[t]
\centering
\includegraphics[width=\linewidth]{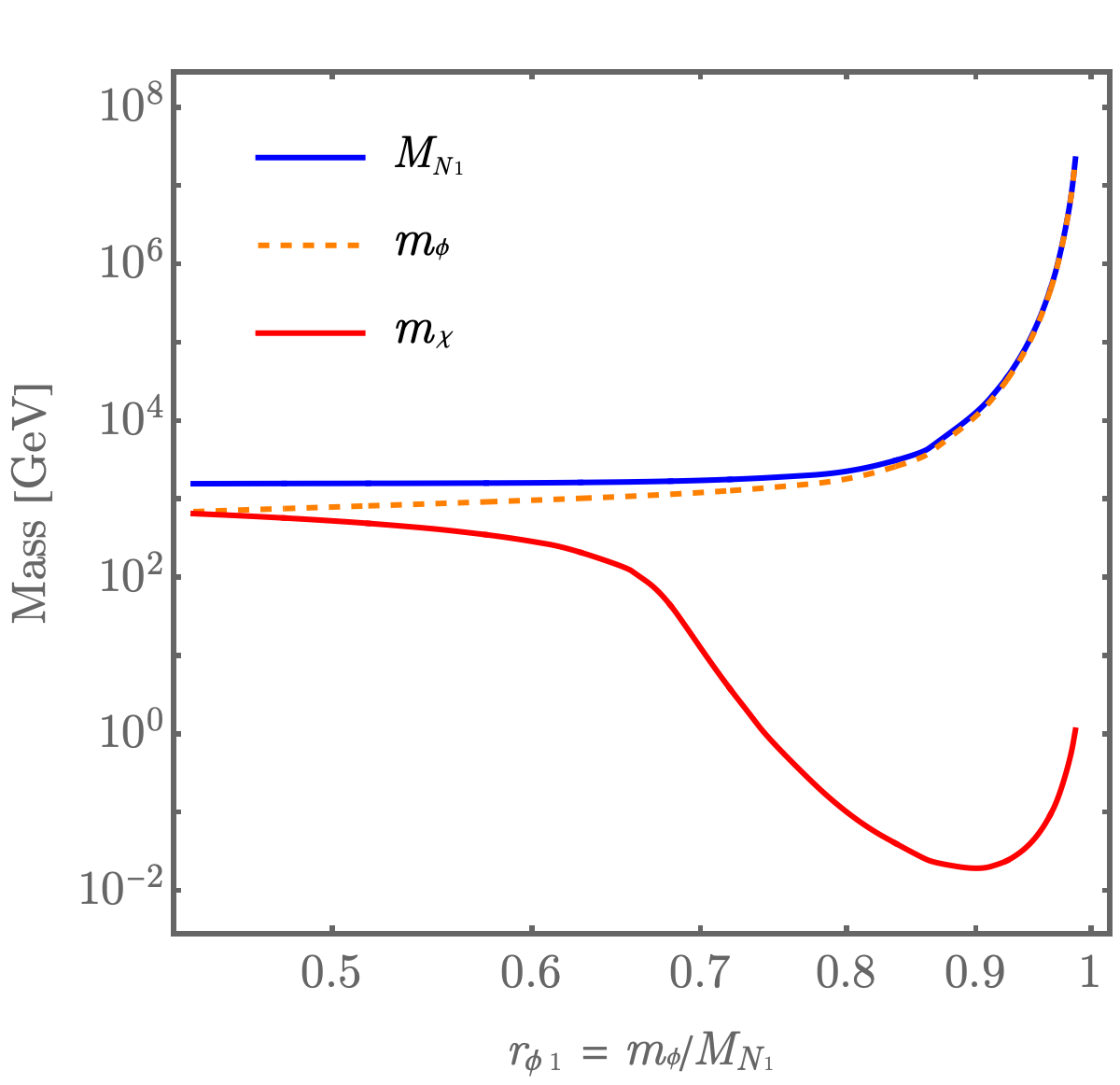} 
\caption{
Dependence of the dark matter mass $m_\chi$ and the required $M_{N_1}$ on the washout-suppression parameter $r_{\phi 1}$
to reproduce the observed asymmetry, showing that increased washout suppression allows lighter RHN masses while maintaining the observed DM relic abundance. The scalar mass $m_\phi$ is shown since DM stability requires $m_\chi<m_\phi$. The remaining parameters are chosen as explained in the text. The curves meet at $r_{\phi 1}\simeq 0.44$, where the required $M_{N_1}$ is lowest.
}
\label{fig:refmass}
\end{figure}

TeV scale leptogenesis is not possible if $r_\phi =m_\phi/M_N \ll 1$ for $M_{N_1} \simeq 2$ TeV due to the strong washout on the dark sector which does not allow for the correct relic abundance as dark matter stability requires $m_\chi < m_\phi$ to be stable. With $m_\phi \lesssim M_{N_1}$, there is significant suppression to the washout in the dark sector. In Fig.~\ref{fig:contour}, the highlighted region shows the parameter points that provide the minimum possible $M_{N_1}$. To determine the possible mass range of the DM for this point, we find the dependence that the required DM mass $m_\chi$ and the required $M_{N_1}$ has on $r_{\phi 1}$. To do this we fix the remaining parameters as $M_{N_2}/M_{N_1} \simeq 1.35$, $|\lambda_1| \simeq \sqrt{M_{N_1}} \cdot 1.5 \times 10^{-9} / \sqrt{[\mathrm{GeV}]}$, $|\lambda_2| = 2 \sqrt{\pi}$, $|y_1| \approx \sqrt{M_{N_1}} \cdot 6.8 \times 10^{-10} / \sqrt{[\mathrm{GeV}]} $, and $|y_2| \approx \sqrt{M_{N_2}} \cdot 6.5 \times 10^{-8} /\sqrt{[\mathrm{GeV}]}$ will provide the required out of equilibrium processes and CP violation, as explained before.
This scenario requires $r_\phi\gtrsim 0.45$ where the phase space suppression due to non zero $m_\phi$ and $m_\chi$ is significant enough to suppress the dark matter washout, which in turn reduces the required value of $m_\chi$ below $m_\phi$. 
\\
The effect of varying $r_{\phi 1}$ has on $M_{N_1}$ and $m_\chi$ for these fixed parameters can be seen in Fig.~\ref{fig:refmass}. For $r_{\phi 1}\gtrsim 0.9$, the phase-space suppression from $r_{\phi 1}$ of the dominant CP violation of the $N_1$ decays (mixing from the loop with $\chi$ and $\phi$ present) significantly reduces overall production thus requiring larger $M_{N_1}$ and larger $m_\chi$ in order to produce the observed baryon asymmetry and correct relic abundance (as $r_{\phi 1} \overset{\sim}{\to} 1$, $\epsilon_\chi,\epsilon_L \overset{\sim}{\to}0$). For smaller $m_\phi$ where $r_{\phi 1}\lesssim 0.6$, the mass of the DM $m_\chi$ is large enough to suppress its own washout, causing an inflection point at $r_{\phi 1}\approx0.6$ such that the $m_\chi$ curve approaches the $m_\phi$ curve more gradually as $r_{\phi 1}$ decreases. The smallest possible $M_{N_1}$ occurs at $r_{\phi 1}\simeq0.44$ where the $m_\chi$ and $m_\phi$ curves meet, since stability requires $m_\chi <m_\phi$ and DM would be unstable for $r_{\phi 1}\lesssim 0.44$.
The DM mass ranges from $m_\chi \approx10^{-2}$ GeV at $r_{\phi\,1}\simeq 0.9$ to $m_\chi\approx10^3$ GeV at $r_{\phi\,1}\simeq 0.44$

\subsection*{Annihilation of the symmetric DM component}

As in any DM scenario with appreciable couplings, a symmetric $\chi\bar\chi$ population is unavoidably produced in the early Universe. To render the observed DM component predominantly asymmetric, one requires an annihilation efficiency of at least
$\langle\sigma v\rangle_{\rm ann}\!\gtrsim\! 3\times10^{-26}\,{\rm cm}^3\,{\rm s}^{-1}$\cite{Falkowski:2011xh, Steigman:2012nb} so that the residual symmetric relic is subdominant. A minimal option is the Higgs portal,
$\mathcal{L}\supset -\lambda_\phi\,H^\dagger H\,\phi^\ast\phi$, which depletes the symmetric part via $s$-channel Higgs exchange. However, even at the Higgs resonance ($m_\chi\simeq m_h/2$), the required $\lambda_\phi$ implies spin-independent scattering rates excluded by direct detection.   Consequently, an efficient \emph{non-Higgs} annihilation channel is required.

A simple and well-studied solution is to introduce a dark photon $A'_D$ with small SM kinetic mixing but sizeable dark coupling, enabling $\chi\bar\chi\to A'_D A'_D$ to deplete the symmetric component without disturbing the asymmetric yield~\cite{Falkowski:2011xh}. An alternative is a light SM-singlet scalar mediator $S$ with a Yukawa coupling $y_S S\bar\chi\chi$, allowing $t$-channel annihilation $\chi\bar\chi\!\to\! SS$. Cosmology then requires that $S$ decays before Big-Bang Nucleosynthesis (BBN). For a Higgs-mixed singlet, $S$ inherits SM-like couplings scaled by $\sin\theta$ and decays promptly enough provided
$\tau_S\!\lesssim\!1\,\mathrm{s}$. Detailed BBN studies of Higgs-portal scalars find $\tau_S\!\lesssim\!0.1\,\mathrm{s}$ throughout most of $2m_\mu\!<\!m_S\!<\!m_h/2$, relaxing to $\mathcal{O}(1\,\mathrm{s})$ only below the di-muon threshold~\cite{Fradette:2017sdd}. 

Using the approximate relation $\Gamma_S\simeq \sin^2\theta\,\Gamma_h^{\rm SM}(m_S)$, one obtains for scalar masses $m_S\sim1\text{--}10~\mathrm{GeV}$ the bound on the mixing parameter of
\[
\sin\theta \;\gtrsim\; \sqrt{\frac{\hbar}{\tau_S\,\Gamma_h^{\rm SM}(m_S)}} 
\;\sim\; \text{few}\times10^{-5}\quad,
\]
consistent with explicit width calculations~\cite{Winkler:2018qyg,Fradette:2017sdd}. This lower bound easily sits \emph{below} collider sensitivities on invisible Higgs decays ${\rm BR}(h\to{\rm inv})$\footnote{For current limits on invisible Higgs decays, see the PDG review and citations therein; e.g. CMS combinations give ${\rm BR}(h\to{\rm inv})\simeq 0.086^{+0.054}_{-0.052}$, and ATLAS/CMS searches set dedicated topology limits at the percent level. See Refs.~\cite{ParticleDataGroup:2024cfk,CMS:2022qva,ATLAS:2023tkt,CMS:2023sdw}.}, and many direct searches for exotic decays, keeping the singlet mediator scenario cosmologically safe yet experimentally elusive at present~\cite{ParticleDataGroup:2024cfk,CMS:2022qva,ATLAS:2023tkt,CMS:2023sdw}. 
In short, with either a dark photon or a light scalar singlet that decays before BBN, the symmetric component can be efficiently erased while preserving the asymmetric production mechanism, rendering the framework effectively indistinguishable from a purely asymmetric DM scenario at current experimental sensitivities.
\\
\section{Observational Prospects}
We consider the impact of a non-zero coupling $\lambda_\phi$ from the portal term in eq.~(\ref{eq:lag}).
In the broken phase, after the Higgs gains a vev, a potential channel for DM direct detection is found assuming that the scalar $\phi$ does not obtain a vev, but does gain a mass contribution from the Higgs.

\begin{figure*}[ht!]
\centering
\includegraphics[width=0.8\textwidth]{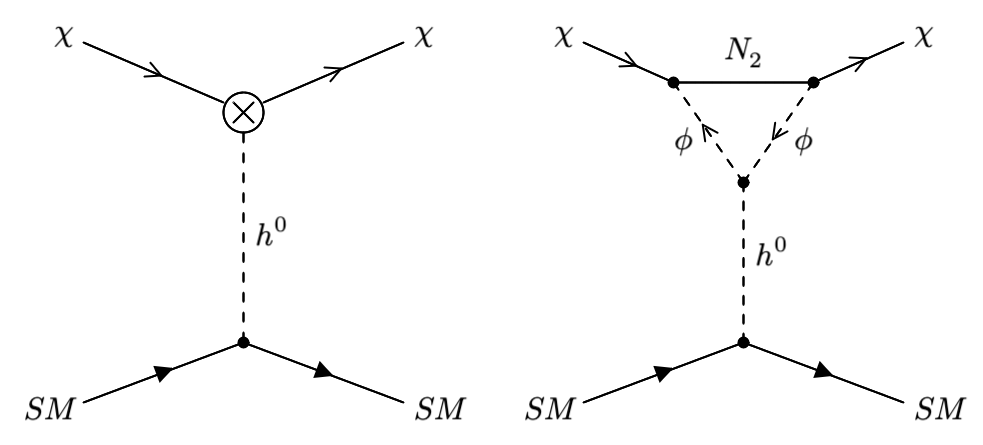} 
\caption{Feynman diagrams for Higgs-mediated dark matter scattering relevant to direct detection: (left) the effective interaction description and (right) the corresponding ultraviolet completion generating the operator at one loop, illustrating the origin of the predicted scattering signal.}
\label{fig:effan}
\end{figure*}

In Fig.~\ref{fig:effan}, the Feynman diagrams for the Higgs-mediated scattering process are shown. The diagram on the left shows the interaction in the effective theory picture, the diagram on the right hand side shows how this interaction arises in the UV-complete theory at the one loop level.
We compute the Wilson coefficient for the induced effective operator $\mathcal{L}\supset C_i \, h^0\,\overline{\chi}\,\chi $ in our theory:
\begin{equation}
    C_i = \frac{|\lambda_2|^2 \lambda_\phi v_{ew}}{16 \pi^2} \frac{M_{N_2}(m_\phi^2-M_{N_2}^2+2M_{N_2}^2 \text{log} [\frac{M_{N_2}}{m_{\phi}}])}{(M_{N_2}^2-m_\phi^2)^2}.
\end{equation}
\\

Using this Wilson coefficient, we find for the Higgs mediated elastic scattering $\chi \,\,SM \rightarrow \chi \, \, SM$, the spin-independent cross section which is given by:
\begin{equation}
    \sigma_{SI}= \frac{C_i^2}{\pi}\left(\frac{m_\chi m_N}{m_\chi +m_N}\right)^2\frac{ \, m_N^2 \, f_N^2}{m_h^4 v^2},
\end{equation}
where the factor $f_N \simeq 0.284$ \cite{Djouadi:2011aa,Ikemoto:2022qxy,Matsumoto:2018acr} encodes the Higgs-nucleon coupling.

\begin{figure}[!htbp]
\centering
\includegraphics[width=\linewidth]{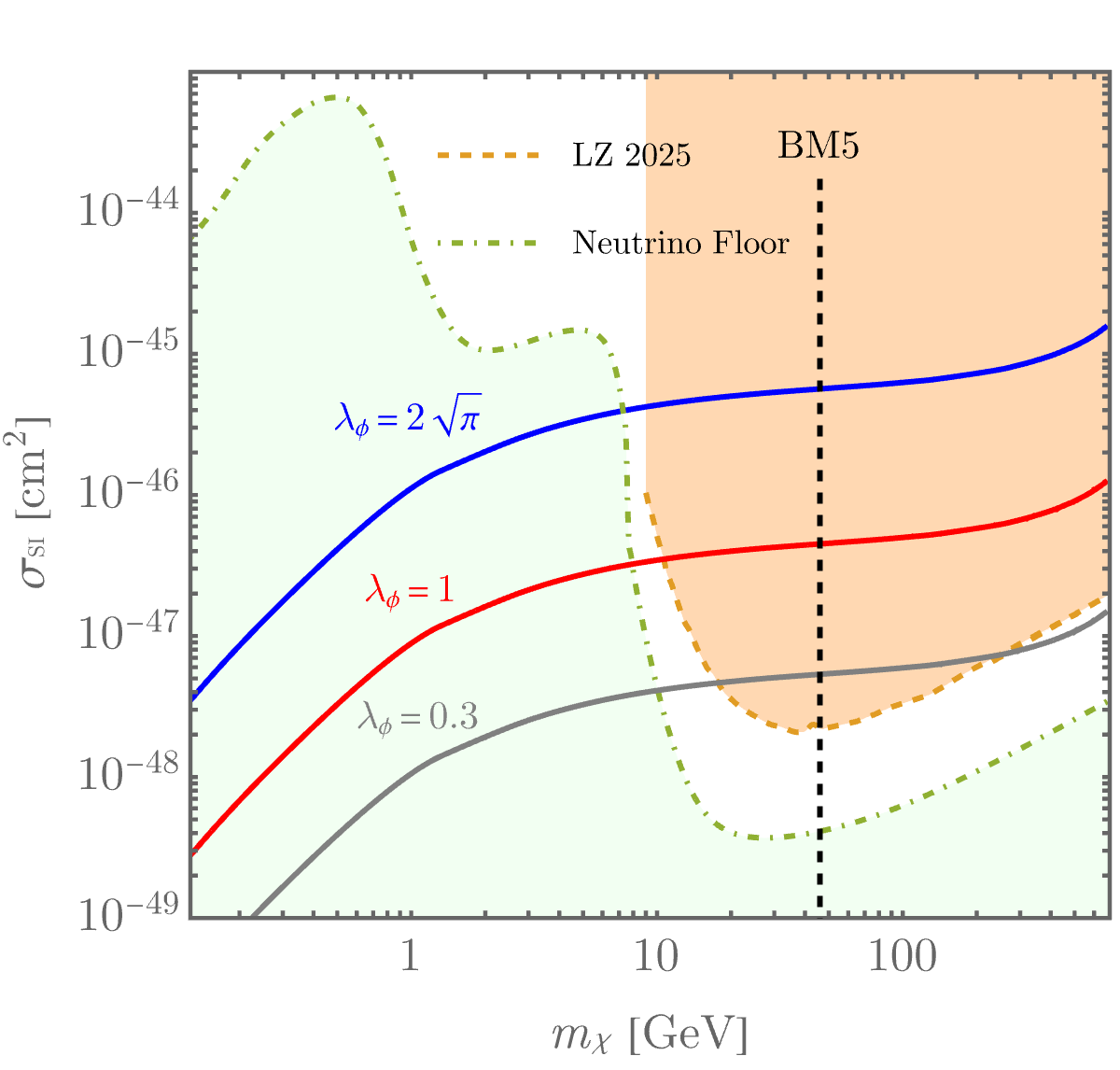} 
\caption{Predicted spin-independent dark-matter–nucleon scattering cross sections as a function of $m_\chi$, where the blue line denotes the perturbative unitarity limit, the orange shaded region indicates current experimental exclusions \cite{LZ:2024zvo}, and the green shaded region shows the neutrino fog, where coherent neutrino scattering limits sensitivity. The remaining parameter space defines the experimentally testable benchmark region of the model with the red and grey lines showing two explicit coupling strength examples.
}
\label{fig:testrange}
\end{figure}
In Fig.~\ref{fig:testrange}, the testable range of parameter space is shown from the computation of the spin-independent cross section directly from Fig.~\ref{fig:refmass}. The shape of the line is dictated by the energy scale needed to produce the correct relic DM abundance. The area below the blue line is the theoretically allowed parameter space based on perturbative unitarity. The shaded orange region has been experimentally ruled out by direct detection searches, and below the orange dashed line lies the parameter space that can be probed with current and near future experimental techniques. 

The green shaded region shows the neutrino fog, the region of parameter space where the coherent scattering from solar neutrinos significantly complicates direct DM detection. Below DM masses of 10 GeV there is significant parameter space within the neutrino fog, and it will be crucially important to explore it in the future.
Current experimental techniques will not be able to penetrate into this region \cite{SENSEI:2019ibb,SENSEI:2020dpa,DAMIC-M:2023gxo,CRESST:2019jnq,CRESST:2022lqw,SuperCDMS:2016wui,SuperCDMS:2023sql}. However, ongoing developments of anisotropic detectors, such as anisotropic crystals~\cite{Blanco:2021hlm,Cook:2024cgm}, point towards a promising direction of probing the DM wind and exploring the theoretically motivated parameter space currently covered by the veil of the neutrino background.  

\begin{figure}[!htbp]
\centering
\includegraphics[width=\linewidth]{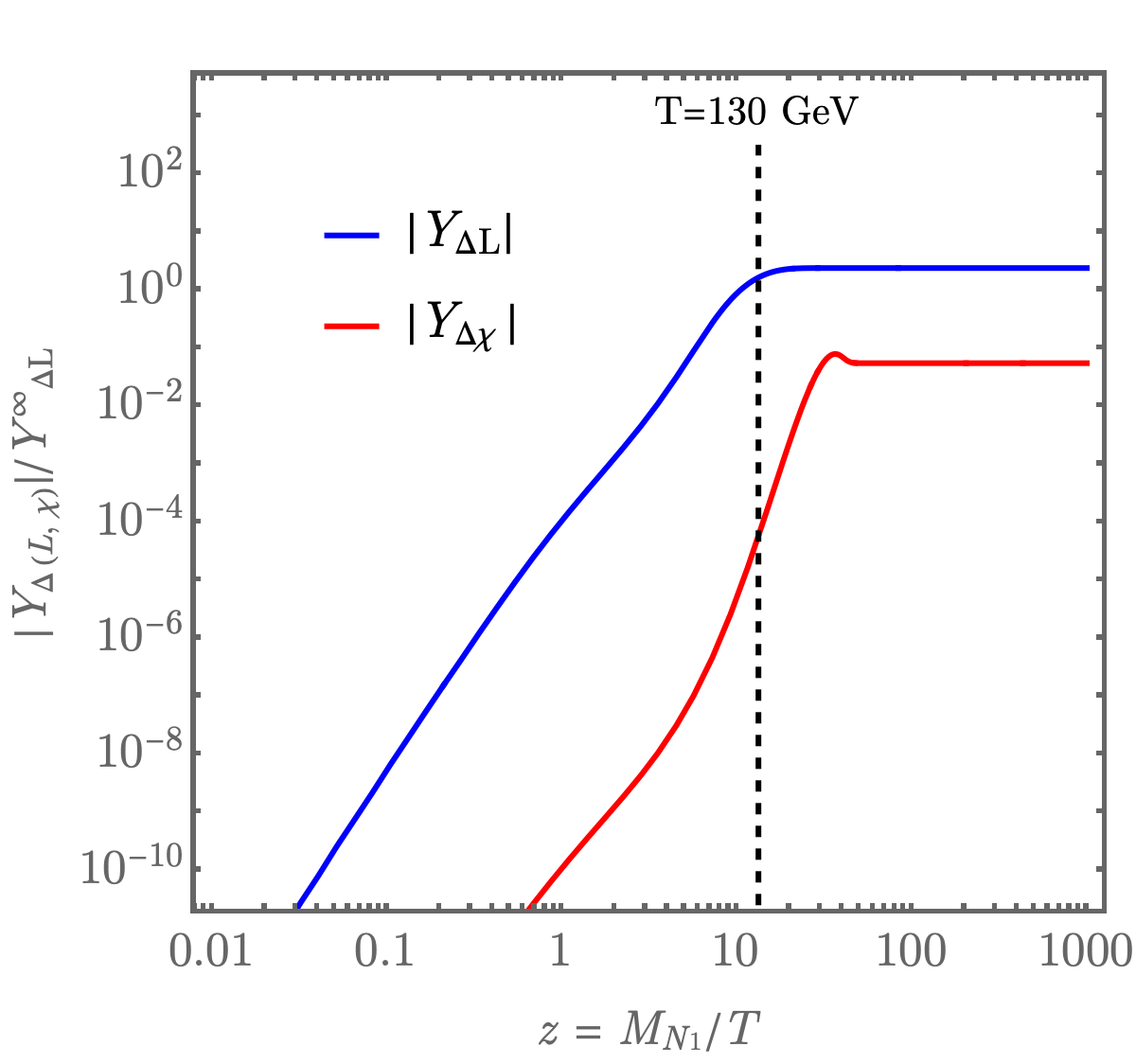}
\caption{Evolution of the lepton and dark asymmetries as a function of $z=M_{N_1}/T$ for benchmark BM5, with the vertical line indicating sphaleron freeze-out, demonstrating successful co-genesis of the baryon and dark matter asymmetries at TeV-scale RHN masses. Finally, the black dashed line shows when sphaleron freeze-out occurs, after which lepton asymmetry is no longer converted to baryon asymmetry}
\label{fig:Result2}
\end{figure}

In Fig.~\ref{fig:Result2} we show the asymmetry generation for BM5 where the DM mass is close to the minimum testable region outside the neutrino fog. The dashed line shows the freeze-out point for the sphaleron process. We emphasise that testable DM signatures in direct detection experiments coincide with relatively low masses of the right handed neutrinos, potentially opening up the prospects of direct experimental confirmation of this coupled matter- and dark matter production mechanism.

\section{Summary}
\label{sec:summary}
Two mechanisms for asymmetric dark matter from leptogenesis have been explored, a wash-in scenario and a co-genesis scenario, in which the decay of a heavy Majorana neutrino simultaneously accounts for light neutrino masses, the baryon asymmetry of the Universe (via sphaleron transfer of a lepton asymmetry), and the dark matter relic abundance explained via an asymmetric dark fermion $\chi$.
CP violation arises from complex Yukawa and dark couplings in the interactions
$N_i \to L H$ and $N_i \to \chi \phi$. From analytic approximations of the Boltzmann equations, which were then validated with numerical calculations, the behaviour of the two mechanisms is understood.

\medskip
\paragraph{(i) Wash-in:}
If the $N_1$ decays to the visible-sector are (approximately) CP-symmetric, a dark asymmetry generated through the $N_1 \to \chi \phi$ decay can be transferred to lepton sector via $N_2$ mediated scatterings. Despite the different origin of the visible asymmetry for this mechanism, successful Baryogenesis reproduces the Davidson-Ibarra bound,
\[
M_{N_1} \gtrsim 10^{9}\ \text{GeV},
\]
up to an overall transfer efficiency factor. 

\medskip
\paragraph{(ii) Co-genesis:}
When CP violation is present in $N_1$ decays to both sectors with a strong hierarchy in the \emph{dark} couplings ($|\lambda_2|\!\gg\!|\lambda_1|\!\simeq\!|y_1|$) but only a mild hierarchy in masses ($M_{N_2}/M_{N_1}\simeq 1.35$), the lepton asymmetry is dominantly produced by $N_1$ while $N_2$ interactions dominate washout. When the mass of the scalar $\phi$ is non-negligible compared to the mass of $N_1$ ($m_\phi\lesssim M_{N_1}$) it suppresses the dark-sector washout, allowing significant survival of the dark asymmetry. Accounting for the sphaleron freeze-out around $T\simeq 130$~GeV \cite{DOnofrio:2014rug}, it was found that successful co-genesis is possible down to
$
M_{N_1} = \mathcal{O}(2~\text{TeV}),
$
well below the DI bound without invoking resonant degeneracy. In this window the dark matter mass spans
$
10^{-2} \ \text{GeV}\lesssim m_\chi\lesssim10^{3}\ \text{GeV},
$
subject to $m_\chi < m_\phi$ for stability.
\medskip
\paragraph{Observational prospects:}
If the interaction $\mathcal{L}\supset - \lambda_\phi\,H^\dagger H \phi^* \phi$ is present, then for DM masses above $m_\chi \gtrsim10$ GeV, there is parameter space that is sensitive to near-future direct detection searches. For DM masses below $m_\chi \lesssim10$ GeV, the parameter space falls into the neutrino fog providing motivation for the development of dedicated tools to probe the neutrino fog.
The low scale for leptogenesis from the co-genesis regime, $M_{N_1}\sim \mathcal{O}(2~\text{TeV})$, could potentially  be probed at future colliders, but the validity of this has yet to be explored. Overall, our results identify a previously unexplored parameter regime in which the cosmological production mechanism directly determines experimentally testable dark-matter interaction benchmarks.

\section{Conclusion}

This work establishes, for the first time, a direct and quantitative connection between mechanisms of asymmetric dark matter generation and the experimental search for dark matter in terrestrial detectors. Within a minimal and testable extension of leptogenesis, the decay of heavy Majorana neutrinos simultaneously accounts for the baryon asymmetry of the Universe, the observed dark matter abundance, and the origin of light neutrino masses. CP violation arises from complex Yukawa and dark-sector couplings in the decays $N_i \to L H$ and $N_i \to \chi \phi$, linking fermion-number violation and dark-sector dynamics within a low-scale type-I seesaw framework. This setup lowers the characteristic scale of lepton-number violation to $\mathcal{O}(\mathrm{TeV})$, addressing the naturalness tension of high-scale seesaw models and opening the possibility of direct experimental probes of leptogenesis.

The framework admits two complementary dynamical regimes: a \emph{wash-in} regime, in which a dark-sector asymmetry is transferred to Standard Model leptons while reproducing the conventional Davidson--Ibarra bound, and a \emph{co-genesis} regime, in which visible and dark asymmetries are generated simultaneously at TeV scales through hierarchical dark couplings and suppressed washout. From a theoretical perspective, the latter scenario is particularly appealing due to the absence of an intermediate heavy mass scale below the Planck scale and the Vissani bound~\cite{Vissani:1997ys}, which would otherwise induce large Higgs mass corrections within the QFT framework. Moreover, the presence of right-handed neutrinos at the TeV scale is independently motivated by fits to electroweak precision observables~\cite{Akhmedov:2013hec}.

In the co-genesis regime, the model predicts a broad dark-matter mass window (see Sec.~\ref{sec:summary}), with a viable low-mass window below a few GeV where dark-sector washout is efficiently reduced by a non-zero scalar mass $m_\phi$. This low-energy realisation of leptogenesis tightly correlates neutrino mass generation, baryogenesis, and dark matter physics, thereby bridging traditionally separate cosmological frontiers within a single predictive framework.

Crucially, the same parameters that govern asymmetric production also determine the strength of dark-matter--nucleon scattering. Through the Higgs-portal interaction $-\lambda_\phi H^\dagger H \phi^\ast \phi$, this model predicts spin-independent scattering cross sections accessible to current and next-generation direct-detection experiments for $m_\chi \gtrsim 10\,\mathrm{GeV}$. Below this threshold, the viable parameter space falls into the so-called \emph{neutrino fog}, where coherent neutrino scattering obscures conventional detection. This strongly motivates the extension of direct-detection sensitivity into the sub-GeV regime with directional or ultra-low-threshold techniques~\cite{SENSEI:2019ibb,SENSEI:2020dpa,DAMIC-M:2023gxo,CRESST:2019jnq,CRESST:2022lqw,SuperCDMS:2016wui,SuperCDMS:2023sql}, which offer a path to overcoming the neutrino background barrier~\cite{Billard:2013qya,Ruppin:2014bra,OHare:2015utx,Bertuzzo:2017tuf}.

More broadly, asymmetric dark matter from leptogenesis emerges as a \emph{physically motivated benchmark scenario} for direct-detection experiments, analogous to the role played by thermal freeze-out in WIMP searches. Rather than treating the dark matter mass and interaction strength as independent parameters, this framework predicts correlated ranges that follow directly from the underlying production mechanism. In this sense, direct-detection experiments represent the primary and most incisive probes of this class of models, providing a unique opportunity to test not only the particle nature of dark matter, but also the cosmological origin of its abundance. We therefore identify direct detection as the key experimental pathway forward for exploring this unified picture of neutrino mass generation, baryogenesis, and asymmetric dark matter.

\vspace{-0.55cm}
\section*{Acknowledgments}
\vspace{-0.45cm}
We thank Jack Shergold and Adam Falkowski for helpful comments on the manuscript. JS acknowledges support from the UK Research and Innovation Future Leader Fellowship~MR/Y018656/1.

\newpage
\begin{widetext}
\section*{Appendix}
\appendix
\noindent
This appendix collects reference formulas used in the main text with short remarks to indicate where each formula appear. Some symbols have been defined again for the first time they appear here.

\section{Casas Ibarra Parametrisation}
The Yukawa couplings of the RHN to the lepton sector are constrained. A standard parametrisation for the constraints on couplings was outlined by Casas and Ibarra in \cite{Casas:2001sr}.
However, that parametrisation does not encompass all possible parameter space. Looking more carefully, there is a larger parameter space to work with. This has been outlined in \cite{Xing:2009vb}.
From \cite{Xing:2009vb} and \cite{Akhmedov:2013hec},
\begin{equation}
    v^2(y^{\dagger} y)\simeq \widehat{M}_{N}^{\,\,T} R^{\dagger}R\,\widehat{M}_N
\end{equation}
where, for the two RHNs case 
\begin{equation}
\widehat{M}_N=
    \begin{pmatrix}
        M_{N_1} & 0 \\ 
        0 & M_{N_2} \\
        0 &0
    \end{pmatrix},
\end{equation}
or for the three RHNs case $\widehat{M}_N= \text{Diag}\{M_{N_1}, \, M_{N_2},\, M_3 \}$, $R$ is defined as
\begin{equation}
    R= -i\cdot U_{PMNS}\,\cdot \widehat{M}_{\nu}^{\,{1}{/2}} \,\cdot \mathcal{O}\,\cdot \widehat{M}_N^{-1/2}
\end{equation}
where $\widehat{M}_\nu= \text{Diag}\{m_1, \, m_2,\, m_3 \}$,
$U_{PMNS}$ is the PMNS matrix, $\mathcal{O}$ is a complex orthogonal matrix defined as $\mathcal{O}=(O_1\cdot O_2\cdot O_3)^\dagger$ where 
\begin{equation}
\begin{alignedat}{3}
O_1&=&\begin{pmatrix}
1&0&0\\
0&\cos\hat\theta_1&-\sin\hat\theta_1\\
0&\sin\hat\theta_1&\cos\hat\theta_1
\end{pmatrix},\qquad
O_2&=&\begin{pmatrix}
\cos\hat\theta_2&0&\sin\hat\theta_2\\
0&1&0\\
-\sin\hat\theta_2&0&\cos\hat\theta_2
\end{pmatrix},\qquad
O_3&=&\begin{pmatrix}
\cos\hat\theta_3&-\sin\hat\theta_3&0\\
\sin\hat\theta_3&\cos\hat\theta_3&0\\
0&0&1
\end{pmatrix},
\end{alignedat}
\end{equation}
satisfying $\mathcal{O}^T \mathcal{O}=\mathbf{1}$ where $\hat{\theta}_i$ is an arbitrary complex angle which may be written as $\hat{\theta}_i=\theta_i+i\,\Theta_i$ where $|\Theta_i|<1$ and $|\theta_i|\leq \pi$.
Normal ordering is assumed with $m_1=0$. The dark couplings are defined as $\lambda_1=e^{i\psi_1}|\lambda_1|$ and $\lambda_2=e^{i\psi_2}|\lambda_2|$.
For BM1-2, the phases used were $\theta_1=-\pi/2, \Theta_1=0.99, \theta_2=\Theta_2=\theta_3=\Theta_3=0$ and the dark phases were $\psi_1=0$ and $\psi_2=-\pi/4$. For BM3-5, the phases used were $\theta_1=-\pi/2, \Theta_1=0.99, \theta_2=-0.01,\Theta_2=0.01,\theta_3=0.015,\Theta_3=0.015$ and the dark phases were $\psi_1=0$ and $\psi_2=-\pi/4$. 
\section{Asymmetry parameters with dark and phi masses}
\noindent
We provide the UV-complete decay width and the CP asymmetries. Denoting $r_{\phi \, i} \equiv m_\phi/M_{N_i}$, $r_{\chi \, i} \equiv m_\chi/M_{N_i}$. The total decay width of $N_i$ is 
\begin{equation}
    \Gamma_{N_{i}} = \frac{(2\,(y_i^{\dagger}y)_{ii}) M_{N_i}}{8 \pi g_N} +( 1-r_{\phi \, i}^2+r_{\chi \, i}^2)\sqrt{\lambda_K(1,r_{\phi \, i}^2,r_{\chi \, i}^2)}\cdot\frac{|\lambda_i|^2\,M_{N_i}}{8 \pi g_N} ,
\end{equation}
where $\lambda_K$ is the Gunnar Källén triangle function given by $\lambda_K(a^2,b^2,c^2)=(a^2-b^2-c^2)^2-4b^2c^2$. 
The asymmetry parameter for the asymmetric decay $N_i\rightarrow LH$ is 
\begin{equation}
\begin{split}
    \epsilon_{L,\,i}&\simeq\sum_j^k\frac{1}{8\pi  }\frac{\text{Br}_{L \,i}}{2(y^{\dagger}y)_{ii}}\left( \frac{M_{N_i}}{M_{N_j}}\text{Im}\left[(y^\dagger y)_{ij}^{2} \right]\right.+
    \\ &
    \left.\frac{M_{N_i} M_{N_j}}{M_{N_j}^2-M_{N_i}^2}\text{Im}\left[2(y^\dagger y)_{ij}^{2}+(y^\dagger y)_{ij}\,\lambda_i^* \lambda_j(1-r_{\phi\,i}^2+r_{\chi \, i}^2) \sqrt{\lambda_K(1,r_{\phi\,i}^2,r_{\chi \, i}^2)}\right]\right).
\end{split}
\end{equation}
For the asymmetric decay $N_i\rightarrow \chi \phi$ the asymmetry parameter is
\begin{equation}
\begin{split}
    \epsilon_{\chi,\,i}&\simeq\sum_j^k\frac{1}{16\pi }\frac{\text{Br}_{\chi \,i}}{|\lambda_i|^2}
    \left( \frac{M_{N_i}}{M_{N_j}}\text{Im}\left[(\lambda_i^* \lambda_j)^2 \right](1-r_{\phi \,i}^2+r_{\chi \, i}^2) \sqrt{\lambda_K(1,r_{\phi \,i}^2,r_{\chi \, i}^2)} \right. 
    \\ &
    \left.+\frac{M_{N_i} M_{N_j}}{M_{N_j}^2-M_{N_i}^2}\text{Im}\left[(\lambda_i^* \lambda_j)^2(1-r_{\phi\,i}^2+r_{\chi \, i}^2) \sqrt{\lambda_K(1,r_{\phi\,i}^2,r_{\chi \, i}^2)}
    +2(y^\dagger y)_{ij}\,\lambda_i^* \lambda_j\right]\right).   
\end{split}
\end{equation}
The simplified asymmetries are given here
\\
\begin{equation}
\begin{aligned}
\epsilon_{L}\ &\simeq\
\frac{M_{N_1}}{M_{N_2}}
\frac{\operatorname{Im}\!\left[3(y^\dagger y)_{12}^{2}+(y^\dagger y)_{12}\,\lambda_1^* \lambda_2\right]}
{8\pi \left(2\,(y^{\dagger}y)_{11}+|\lambda_1|^2\right)},
\qquad\qquad
\epsilon_{\chi}\ &\simeq\
\frac{M_{N_1}}{M_{N_2}}
\frac{\operatorname{Im}\!\left[(\lambda_1^* \lambda_2)^2+(y^\dagger y)_{12}\,\lambda_1^* \lambda_2\right]}
{8\pi \left(2(y^{\dagger}y)_{11}+|\lambda_1|^2\right)},
\end{aligned}
\label{eq:epsilons}
\end{equation}
\\
which correspond to the limit $r_{\phi\,i},r_{\chi \, i}\to 0$.

\section{Analytical Transfer}
\noindent
The following expression gives the lepton-number conserving transfer kernel $I_{T_-}$ that appears in eq.~(\ref{eq:dlbltz}). It is obtained by Taylor expanding the exact $2 \leftrightarrow 2$ cross section in $r_{\phi \,i}=m_\phi/M_{N_i}$ and then thermally averaging it (as was done for $I_{T_+}$):
\begin{equation}
    I_{T_{-,\,i}}(z_i)=\frac{1}{\pi}\frac{\Gamma_{N_i}}{M_{N_i}}\frac{4 r_{\phi \, i} \left(r_{\phi \, i}^4-2r_{\phi \, i}^2+2\right) \left(r_{\phi \, i} z_i \left(r_{\phi \, i}^2 z_i^2+24\right) K_0(r_{\phi \, i} z_i)+8 \left(r_{\phi \, i}^2 z_i^2+6\right) K_1(r_{\phi \, i}
   z_i)\right)}{\left(r_{\phi \, i}^4-1\right)^2 z_i^3}.
\end{equation}

\section{cross section functions}
\noindent
The thermally averaged cross section functions in eq.~(\ref{eq:thermcross}) ($I_W$, $I_{T_+}$, $I_{T_-}$) are built from the following $2 \leftrightarrow 2$ cross sections; we define $\hat{s}\equiv s/M_{N_i}^2$ and $\hat{\Gamma}_i\equiv \Gamma_{N_i}/M_{N_i}$. We also define $\overline{\Gamma}_i=\hat{\Gamma}_i$ when $r_\phi,r_\chi=0$.
The cross section functions are defined as 
\begin{equation}
    f_{x,\,i}(\hat{s}) = \frac{1}{8\pi \,\overline{\Gamma}_i^2} \frac{\sigma_{ab}(\hat{s})\,\,p_{cm}^{\,2}(\hat{s}) }{\widehat{\text{Br}}_a \widehat{\text{Br}}_b}, 
\end{equation}
where $p_{cm}$ is the centre of mass momentum, $\widehat{\text{Br}}_i={\text{Br}}_i$ when $r_\phi,r_\chi=0$, and $\sigma$ is the cross section. The cross section have been checked with Ref.~\cite{Giudice:2003jh}.
For washout for processes of the form ${L}\,H \longleftrightarrow \overset{\text{\scriptsize\textbf{—}}}{L}\,H^{\dagger} $
with lepton-number violation:
\begin{equation}
\begin{split}
    f_{W_1 ,\,i}(\hat{s})&= \frac{\hat{s}}{2\,\,((\hat{s}-1)^2+\hat{\Gamma}_i^2)}+ \frac{\hat{s}-\log(\hat{s})}{\hat{s}} - \frac{(\hat{s}-1)}{(\hat{s}-1)^2+\hat{\Gamma}_i^2}\cdot \left(\frac{(\hat{s}+1)\log (\hat{s}+1)-\hat{s}}{\hat{s}}\right),\\
\end{split}
\end{equation}
and for washout for processes of the form ${L}\,{L} \longleftrightarrow H^{\dagger} H^{\dagger}$ with lepton-number violation:
\begin{equation}
\begin{split}
    f_{W_2,\,i}(\hat{s})=& \frac{\hat{s}}{2(\hat{s}+1)}+\frac{\log(\hat{s}+1)}{2(\hat{s}+2)}.
\end{split}
\end{equation}
\\
\\
For washout for processes of the form 
${\chi}\,\phi \longleftrightarrow \overset{\text{\scriptsize\textbf{—}}}{\chi}\phi^{*} $
with lepton-number violation:

\begin{equation}
\begin{split}
    f_{W_1 ,\,i}(\hat{s})&= \frac{(\hat{s} -r_{\phi \, i}^2+r_{\chi \, i}^2)^2(\lambda_K(\hat{s},r_{\phi \, i}^2,r_{\chi \, i}^2))}{2\hat{s}^3\,\,((\hat{s}-1)^2+\hat{\Gamma}_i^2)}\\
    &+ \frac{1}{\hat{s}}\left(  \frac{(\hat{s}+1-2r_{\phi \, i}^2)(\lambda_K(\hat{s},r_{\phi \, i}^2,r_{\chi \, i}^2))}{(\hat{s}-(r_{\phi \, i}^2-r_{\chi \, i}^2)^2)(\hat{s}+1-2r_{\phi \, i}^2-2r_{\chi \, i}^2)}   
    -\log \left(  \frac{\hat{s}(\hat{s}+1-2r_{\phi \, i}^2-2r_{\chi \, i}^2)}{\hat{s}-(r_{\phi \, i}^2-r_{\chi \, i}^2)^2}   \right)      \right) \\
    &+ \frac{2(\hat{s}-1)}{\hat{s}^2\,\,((\hat{s}-1)^2+\hat{\Gamma}_i^2)} \left( \lambda_K(\hat{s},r_{\phi \, i}^2,r_{\chi \, i}^2)  -\hat{s}(\hat{s}+1-2r_{\phi \, i}^2) \log \left(  \frac{\hat{s}(\hat{s}+1-2r_{\phi\,i}^2-2r_{\chi\,i}^2)}{\hat{s}-(r_{\phi \, i}^2-r_{\chi \, i}^2)^2}   \right)\right),\\
\end{split}
\end{equation}
\noindent and for washout for processes of the form ${\chi}\,{\chi} \longleftrightarrow \phi^{*} \phi^{*}$ 
with lepton-number violation: 
\begin{equation}
\begin{split}
    f_{W_2,\,i}(\hat{s})=& \,\frac{\sqrt{\hat{s}-4r_{\phi \, i}^2}\sqrt{\hat{s}-4 r_{\chi \, i}^2}(\hat{s}-2r_{\chi \, i}^2)}{2 \hat{s} (\hat{s}+1-2r_{\phi \, i}^2-2r_{\chi \, i}^2+(r_{\phi \, i}^2-r_{\chi \, i}^2)^2)}\\
    &+\frac{\hat{s}-2r_{\chi \, i}^2}{ \hat{s}(\hat{s}+2(1-r_{\phi \, i}^2-r_{\chi \, i}^2))} \log \Biggl[  \frac{\hat{s}+\sqrt{\hat{s}-4 r_{\phi \, i}^2}\sqrt{\hat{s}-4 r_{\chi \, i}^2}+2(1-r_{\phi \, i}^2-r_{\chi \, i}^2)}{\hat{s}-\sqrt{\hat{s}-4 r_{\phi \, i}^2}\sqrt{\hat{s}-4 r_{\chi \, i}^2}+2(1-r_{\phi \, i}^2-r_{\chi \, i}^2)}\Biggr].
\end{split}
\end{equation}

\noindent For lepton-violating transfer s-channel of the form 
${\chi} \,\phi \longleftrightarrow \overset{\text{\scriptsize\textbf{—}}}{L}\,H^{\dagger} $
:
\begin{equation}
\begin{split}
    f_{T_+\,s,\,i}(\hat{s})=& \frac{(\hat{s}-r_{\phi \, i}^2+r_{\chi \, i}^2)\sqrt{\lambda_K(\hat{s},r_{\phi \, i}^2,r_{\chi \, i}^2)}}{2 \hat{s} ((\hat{s}-1)^2+\hat{\Gamma}_i^2)},
\end{split}
\end{equation}
and for lepton-violating transfer t and u channels of the form ${\chi}\,{L} \longleftrightarrow\phi^{*}H^{\dagger} $ and ${\chi}\,H\longleftrightarrow \overset{\text{\scriptsize\textbf{—}}}{L}\,\phi^{*} $:
\begin{equation}
\begin{split}
    f_{T_+\,t,\,i}(\hat{s})=& \frac{(\hat{s}-r_{\phi \, i}^2)(\hat{s}-r_{\chi \, i}^2)^2}{\hat{s}(\hat{s}^2+\hat{s}(1-r_{\phi \, i}^2-r_{\chi \, i}^2)+r_{\phi \, i}^2 r_{\chi \, i}^2)} \\
    &+ \frac{1}{\hat{s}}\Biggl(\frac{(\hat{s}-r_{\phi \, i}^2)(\hat{s}-r_{\chi \, i}^2)(\hat{s}+1-r_{\phi \, i}^2)}{\hat{s}^2+\hat{s}(1-r_{\phi \,i}^2 -r_{\chi \,i}^2 )+r_{\phi \,i}^2r_{\chi \,i}^2}  -\log\left[\frac{\hat{s}^2+\hat{s}(1-r_{\phi \,i}^2 -r_{\chi \,i}^2 )+r_{\phi \,i}^2r_{\chi \,i}^2}{\hat{s}} \right]\Bigg).
\end{split}
\end{equation}
For lepton-conserving transfer s-channel of the form ${\chi}\,\phi \longleftrightarrow {L}\,H $:
\begin{equation}
\begin{split}
    f_{T_-\,s,\,i}(\hat{s})=& \frac{(\hat{s}-r_{\phi \, i}^2+r_{\chi \, i}^2)\sqrt{\lambda_K(\hat{s},r_{\phi \, i}^2,r_{\chi \, i}^2)}}{2 ((\hat{s}-1)^2+\hat{\Gamma}_i^{\,2})}, 
\end{split}
\end{equation}
and for lepton-conserving transfer t and u channels of the form ${\chi}\overset{\text{\scriptsize\textbf{—}}}{L}\longleftrightarrow \phi^{*}H $ and ${\chi}\,H^{\dagger}\longleftrightarrow {L} \ \phi^{*}$ :
\begin{equation}
\begin{split}
    f_{T_-\,t,\,i}(\hat{s})=& \frac{1}{\hat{s}^2}\Biggl(\hat{s}\,(\hat{s} - r_{\phi \, i}^{2} + 2)\,\log\!\left(
    \frac{\hat{s}^2+\hat{s}(1-r_{\phi \, i}^2-r_{\chi \, i}^2)+r_{\phi \, i}^2r_{\chi \, i}^2}{\hat{s}}
    \right)\\
    &-
    \frac{(\hat{s} - r_{\phi \, i}^{2})(\hat{s} - r_{\chi \, i}^{2})\left(2 \hat{s}^2 +\hat{s}(2-2r_{\phi \, i}^2-r_{\chi \, i}^2)+r_{\phi \, i}^2 r_{\chi \, i}^2\right)}
    {\hat{s}^{2} + \hat{s}\,(1-r_{\phi \, i}^{2} - r_{\chi \, i}^{2} ) + r_{\phi \, i}^{2} r_{\chi \, i}^{2}}\Biggr) \\
    &+\frac{(\hat{s}-r_{\chi \, i}^2)}{\hat{s}}\Biggl( \log\left[\frac{\hat{s}^2+\hat{s}(1-r_{\phi \,i}^2 -r_{\chi \,i}^2 )+r_{\phi \,i}^2r_{\chi \,i}^2}{\hat{s}} \right] -\frac{(\hat{s}-r_{\phi \,i}^2)(\hat{s}-r_{\chi \,i}^2)}{\hat{s}^2+\hat{s}(1-r_{\phi \,i}^2 -r_{\chi \,i}^2 )+r_{\phi \,i}^2r_{\chi \,i}^2}\Biggl). 
\end{split}
\end{equation}

\section{Definitions \label{sec:apdef}}
\noindent
For reference, the equilibrium yield in eq.~(\ref{eq:deptboltz}) is defined as \cite{Edsjo:1997bg}:
\begin{equation}
\label{eq:YNeq}
    Y_{N_i}^{eq}=\frac{g_{N_i}M_{N_i}^3}{s(T=M_{N_i})}\frac{1}{2\pi^2}z_i^2 K_2(z_i),
\end{equation}
where $s(T)$ is entropy density,
and the normalisation entering eq.~(\ref{eq:YNeqap}) such that the analytic approximation agrees with the above definition when $z=z_0$ is:
\begin{equation}
\label{eq:approxconst}
    A(z_0) = \frac{Y_{N_i}^{eq}(z_0)}{z_0^{3/2}e^{-z_0}}=\frac{45\,g_{N_i}}{(2\pi^2)^2 g_\star}z_0^{1/2}\,e^{z_0}\,K_2(z_0).
\end{equation}
    
\end{widetext}
\newpage
\bibliographystyle{JHEP}
\bibliography{reference.bib}

\end{document}